\documentclass[prd, 10pt,aps,nofootinbib, floatfix, notitlepage,twocolumn,longbibliography,superscriptaddress]{revtex4-1}

\usepackage[utf8]{inputenc}
\usepackage[dvips]{graphicx} 
\usepackage{amsmath,amsfonts,amssymb,slashed,float,bm,algorithmic}
\usepackage{bbold,wasysym}
\usepackage{array,multirow}
\usepackage[usenames,dvipsnames]{xcolor} 
\usepackage{soul}
\usepackage{hyperref}
\DeclareUnicodeCharacter{2212}{-}

\definecolor{blop}{RGB}{110, 180, 181}

\definecolor{nicer_blue}{RGB}{28, 96, 214}
\hypersetup{
	colorlinks=true,
	urlcolor=nicer_blue,
	citecolor=nicer_blue,
}

\newcommand{\LCDM}{$\Lambda$CDM}

\newcommand{\fede}{f_{\rm ede}}

\begin{document}
\title{Procoli: Profiles of cosmological likelihoods}

\author{Tanvi Karwal}
\email{karwal@uchicago.edu}
\affiliation{Kavli Institute for Cosmological Physics, the University of Chicago, IL 60637, USA}
\author{Yashvi Patel}
\affiliation{Department of Physics and Astronomy, 
Swarthmore College, Swarthmore, PA 19081, USA}
\author{Alexa Bartlett}
\affiliation{Department of Physics, University of California, Berkeley, CA 94720, USA}
\author{Vivian Poulin}
\affiliation{Laboratoire Univers \& Particules de Montpellier (LUPM),
CNRS \& Universit\'e de Montpellier (UMR-5299),
Place Eug\`ene Bataillon, F-34095 Montpellier Cedex 05, France}
\author{Tristan L. Smith}
\affiliation{Department of Physics and Astronomy, 
Swarthmore College, Swarthmore, PA 19081, USA}
\author{Daniel N. Pfeffer}
\affiliation{Independent Researcher}

\begin{abstract}
Frequentist profile likelihoods have seen a resurgence in cosmology, offering an alternative to Bayesian methods as they can circumvent the impact of prior-volume effects. 
This paper presents Procoli, a fast and accessible package to obtain profile likelihoods in cosmology, available on GitHub and PyPI. 
Procoli seamlessly integrates with MontePython, incorporating all its available data likelihoods, as well as any modified versions of CLASS. 
This paper provides a comprehensive overview of the Procoli code, detailing the simulated-annealing optimizer at its core and the sequential computation of the profile. 
An an example, we use the early dark energy model which is afflicted by prior-volume effects to illustrate the code's features. 
We validate its optimizer with mock data, and compare optimization techniques for both the global minimum and the profile. 
Procoli further enables splitting profiles into their component contributions from individual experiments, offering nuanced insights into the data and model.
As a valuable addition to the cosmologist's toolkit, Procoli supplements existing Bayesian codes, contributing to more robust parameter constraints in cosmological studies.
\end{abstract}
 
\maketitle

\section{Introduction} \label{Sec:Intro}

Over the last several years, cosmology has relied on Bayesian statistics for constraining models and parameters with data \cite{Lewis:2002ah, Cousins:1994yw}. 
This approach explores the data likelihood $\mathcal{L}$ over model parameter space via Bayes theorem, determining the parameter posterior distributions $\mathcal{P}$ 
\begin{equation}
    \mathcal{P}(\bm{\theta}|\bm{d}) = 
    \frac{ \Pi(\bm{\theta}) \mathcal{L}( \bm{d} |\bm{\theta})} 
    {\mathcal{E}(\bm{d})} \,,
    \label{eq:bayes_theorem}
\end{equation}
as a product of the prior probability distribution $\Pi$ of input model parameters $\bm{\theta}$ and the likelihood $\mathcal{L}$ of the data $\bm{d}$ given the model parameters. 
The evidence $\mathcal{E}(\bm{d})$ of the observations is usually treated as a normalization constant and ignored. 
Any posteriors $\mathcal{P}$ obtained from Bayesian methods then implicitly depend on the priors $\Pi$, making this a powerful approach to incorporate expectations for and prior knowledge of $\bm{\theta}$, but they may not encapsulate ignorance of $\bm{\theta}$ well (see Jeffreys priors \cite{Jeffreys:1946} for an attempt). 
Even flat priors on $\bm{\theta}$ can be subjective and informative, as they are flat priors in a chosen basis \cite{Feldman:1997qc, Hill:2020osr}. 
They are hence non-uniform and indeed informative priors in a different basis, or a different parameterization. 

For models that have priors that do not strongly influence the posteriors, posteriors resemble the data likelihood itself. 
This for example is the case for the standard \LCDM\ model of cosmology, whose parameters have Gaussian posteriors that are well-constrained within broad, uninformative, flat priors, across multiple parameterizations (eg. using $\ln 10^{10} A_s$ vs $A_s$) and the Bayesian and frequentist approaches agree \cite{Planck:2013nga}. 

The phenomenological \LCDM\ model has provided a good fit to numerous data sets across several redshifts and physical scales \cite{Planck:2018vyg, BOSS:2016wmc,Pan-STARRS1:2017jku} until recently. 
Lately, notable tensions and anomalies have emerged between different cosmological data sets, including the Hubble and weak-lensing $S_8$ tensions \cite{Verde:2019ivm,Abdalla:2022yfr,Riess:2023egm, Secco:2022kqg,DES:2021bvc,DES:2021vln,Li:2023azi,Dalal:2023olq}. 
This, coupled with the phenomenological nature of \LCDM\, motivates the search for physics beyond the standard \LCDM\ model. 

Such models are often parameterized by capturing the departure from \LCDM\ via a single parameter ($f$ below), with additional parameters describing the properties of the newly added physics. 
Ultimately this means that as $f$ goes to its \LCDM\ limit, the prior volume balloons because varying the other new-physics parameters no longer impacts observations and leaves the likelihood unchanged. This leads to a posterior distribution which has a strong dependence on the choice of priors \cite{Smith:2020rxx,Herold:2021ksg,Planck:2013nga}. 

Bayesian statistics are known to be susceptible to prior-volume effects \cite{Herold:2021ksg, Hamann:2011hu}. 
As more beyond-\LCDM\ models are explored, it is prudent to develop alternative techniques to constrain models that specifically mitigate prior-volume effects. 
One such method is profile likelihoods, a frequentist approach with a long history of use in cosmology as well as other fields including experimental particle physics \cite{Planck:2013nga,Hamann:2011hu,Palanque-Delabrouille:2019iyz,Reeves:2022aoi,Feldman:1997qc,Ranucci:2012ed}. 
Frequentist statistics do not rely on priors, eliminating prior-volume effects in this framework. 
While both approaches provide parameter constraints, frequentist confidence intervals fundamentally differ from Bayesian credible intervals \cite{Feldman:1997qc}. 
Results from a profile likelihood are statements about the likelihood $\mathcal{L}(\bm{d}|\bm{\theta})$, while Bayesian intervals are statements about the posterior $\mathcal{P}(\bm{\theta}|\bm{d})$. 
A 1D profile likelihood on the parameter $\theta_i$ directly traces $\mathcal{L}$ and maximizes it at each point $\theta_i'$ in parameter space as 
\begin{equation}
    \mathcal{L}(\theta_i) = \max \mathcal{L}(\bm{\theta}|\theta_i=\theta_i') \,, 
\end{equation}
where the maximization is over all parameters $\theta_{j \neq i} \in \bm{\theta}$. 
This can then be generalized to higher dimensions for 2D, 3D and so on profiles of the likelihood. 
A notable advantage of this approach is the invariance of profile likelihoods under parameter redefinitions, as the likelihood remains constant regardless of whether a linear or logarithmic parameter is used (eg. $A_s$ vs. $\ln 10^{10} A_s$), or even if an alternative parameter is used (eg. $\theta_*$ vs. $H_0$). 
The result is a curve of the effective $\chi^2 = -2\log\mathcal{L}$ of the data vs. the profile likelihood parameter $\theta_i$. 
Parameter confidence intervals can be built from this using various constructions that become equivalent in the typical case of Gaussian measurement errors far away from the physical boundaries of a parameter. 
The Neyman construction \cite{Neyman:1937uhy}, for a single parameter, is the most straightforward: a $\Delta \chi^2=1$ relative to the global minimum $\chi^2_{\min}$ denotes the $1\sigma$ (68\%) region, $\Delta \chi^2 = 4$ is the $2\sigma$ (95\%) and so on. 
Interestingly, this construction does not rely on the profile itself actually being Gaussian (i.e. a parabolic $\chi^2$ curve). 
It is enough that there exists a parameter redefinition in which the profile becomes Gaussian. 
For the remainder of this paper, we will assume this to be the case.
It can also be easily generalized for N-dimensional profiles by computing the (complementary of the) cumulative of the $\chi^2$ distribution $P$ for $N_{\rm d.o.f}$, $P_{\rm N_{d.o.f.}}(\chi^2 \geq \Delta \chi^2)$. 
For more complicated cases, one may rely alternatively on the Feldman-Cousins construction \cite{Feldman:1997qc}. 
See Refs.~\cite{Planck:2013nga,Herold:2021ksg} for a concrete example in cosmology.

Over the last few years, profile likelihoods have regained favor in cosmology, owing to the popularity of several beyond-\LCDM\ models that experience prior-volume effects, 
including models of neutrinos \cite{Planck:2013nga,Reeves:2022aoi,Palanque-Delabrouille:2019iyz,Hamann:2011hu}, 
dark matter \cite{Holm:2023uwa,Nygaard:2022wri,Li:2022mdj, Holm:2022kkd}, 
and early dark energy \cite{Smith:2020rxx,Herold:2021ksg}, several of these proposed as solutions to the Hubble tension. 
Investigating these models via Bayesian techniques can bias results, but deploying profile likelihoods can become computationally expensive. 
In this methodology paper, we present a quick and easy publicly available tool to obtain profile likelihoods in cosmology called Procoli\footnote{Procoli is available via \href{https://github.com/tkarwal/procoli}{GitHub} as well as \href{https://pypi.org/project/procoli/}{PyPI}}. 
Procoli wraps MontePython \cite{Audren:2012wb,Brinckmann:2018cvx}, a public sampler for cosmological data, automatically including all the data likelihoods available with MontePython. 
The cosmological back-end for calculating observables is assumed to be the CLASS Boltzmann code \cite{Blas:2011rf}. 
Although we present this tool wrapping MontePython, the same principles can also be immediately applied to Cobaya \cite{Torrado:2020dgo,2019ascl.soft10019T}, another public sampler in cosmology that wraps both the two most used Boltzmann codes CAMB \cite{Howlett:2012mh,Lewis:1999bs} and CLASS. 
The Cobaya version of our code is still in development, but available for use upon request. 
We also welcome collaboration on this or any other front of further developing Procoli. 

This paper is structured as follows. 
In Section \ref{sec:code}, we describe the Procoli code. 
Specifically, we describe our simulated-annealing optimizer at the heart of the profile-likelihood approach in \ref{subsec:optimizer},
the sequential progression of this optimizer to calculate the profile in \ref{subsec:initialization}, 
and how to run Procoli and its parameters and inputs in \ref{subsec:implementation}. 
Then, using early dark energy and \LCDM\ as examples, we show comparisons and features of the code in Section \ref{sec:examples}, 
including validating the global optimizer in \ref{subsec:min_validation}, 
comparing simulated annealing to a deterministic quasi-Newton optimizer in \ref{subsec:compare_mcmc_minuit}, 
and splitting profiles over the experiments they fit to in \ref{subsec:chi2_per_exp}. 
Finally, we summarize and look ahead in Section \ref{sec:conclusions}.

\section{The Procoli code}
\label{sec:code}

\subsection{Optimizer}
\label{subsec:optimizer}

Profile likelihoods are principally based on optimizers. 
At every point in parameter space, or for our purposes, at every point of parameter $\theta_i$ for which a profile is desired, the likelihood function $\mathcal{L}$ must be optimized over all other parameters $\theta_{j \neq i} \in \bm{\theta}$. 
Hence, an efficient and accurate optimizer is at the core of our profile likelihood code. 
We employ a simulated-annealing algorithm \cite{Hannestad:2000wx} to find the maximum likelihood at each point, a scheme that is rapidly becoming the industry standard for profile-likelihood studies in cosmology \cite{Schoneberg:2021qvd,Herold:2021ksg,Holm:2023uwa,Goldstein:2023gnw,Reeves:2022aoi}. 

Simulated annealing (SA) is a stochastic method that modifies a Metropolis-Hastings (MH) MCMC algorithm by introducing a ``temperature'' $T$ to the likelihood, and changing the step size $\delta$ of the sampler in accordance \cite{Hannestad:2000wx}. 
Conventionally, at low temperatures, any gradients in the likelihood are enhanced, encouraging the stochastic MCMC algorithm to converge to the maximum likelihood. 
Practically, this works as follows. 
MCMCs \cite{Lewis:2002ah} randomly take ``steps'' in parameter space, that is, they perturb the input parameters, calculate the observables, compare those to data, and either accept this new step, moving their position, or reject the new step, remaining at the current position in parameter space. 
Vanilla MH is the algorithm underlying most MCMC codes \cite{Audren:2012wb, Brinckmann:2018cvx, Torrado:2020dgo}, and it decides whether to accept or reject the step based on their likelihoods as follows:
\begin{algorithmic}
  \IF{ $\mathcal{L_{\rm new}} \geq \mathcal {L_{\rm old}}$ }
    \STATE accept
  \ELSE
    \STATE draw a random number $x \in [0,1]$ from a uniform distribution
    \IF{$x < \mathcal{L_{\rm new}} / \mathcal {L_{\rm old}}$}
        \STATE accept 
    \ELSE 
        \STATE reject
    \ENDIF
  \ENDIF
\end{algorithmic}
Modifying this, SA tempers the acceptance algorithm by exponentiating the likelihood ratio in the case of $\mathcal{L_{\rm new}} / \mathcal {L_{\rm old}} < 1$ usually with a temperature $T < 1$: 
\begin{algorithmic}
\IF{$x < (\mathcal{L_{\rm new}} / \mathcal {L_{\rm old}})^{1/T}$}
    \STATE accept 
\ELSE 
    \STATE reject
\ENDIF
\end{algorithmic}
For MH, $T=1$. 
As $T$ is decreased, features in the likelihood are sharpened, such that the probability to reject points that lower the likelihood increases with smaller $T$. 

Besides tempering the likelihood, SA also modifies the step size or jumping factor $\delta$ of the sampler, which dictates how large a step $\Delta \theta_j$ is taken in each parameter per iteration in conjunction with the covariance matrix $\Sigma$ as 
\begin{equation}
    \Delta \theta_j \simeq \delta \sqrt{\Sigma_{jj}} \,.
\end{equation} 
For MH, the recommended jumping factor hovers around $\delta = 2.4$ \cite{Dunkley:2004sv} (unless this is adaptively updated \cite{Brinckmann:2018cvx}). 
However, assuming we are close to the maximum likelihood, as $T$ decreases and the likelihood is sharpened, a large step size will drastically reduce the acceptance rate. 
In this scenario, most proposals will be rejected as they worsen the current likelihood, stepping too far away from the maximum likelihood. 
Therefore, as the temperature of the likelihood is decreased, the jumping factor must also decrease. 
Usually, SA algorithms attempt to vary $T$ and $\delta$ such that the acceptance rate is roughly constant \cite{Schoneberg:2021qvd}.

In Procoli, we provide empirical recommendations for $T$ and $\delta$ that balance efficiency and accuracy, but these can also be input by the user. 
Ultimately, the progression of $T$ and $\delta$ can be improved by an adaptive algorithm which we leave to future updates to the code. 

We find that an SA optimizer implemented as described above substantially improves over Minuit\footnote{
\href{https://iminuit.readthedocs.io/en/v1.5.4/index.html\#}{Minuit} is publicly available through PyPI. 
} \cite{dembinski_2023_8249703,James:1975dr}, a quasi-Newtonian optimizer often used with MontePython \cite{Planck:2013nga}, which is deterministic in nature. 
The stochastic approach of SA can help the optimizer escape local minima, leading to more robust outcomes as shown in Sec. \ref{subsec:compare_mcmc_minuit}. 
An additional advantage of this approach is that it wraps an MCMC, for which one usually activates multiple parallel chains at once. 
Under SA, each chain effectively provides a semi-independent optimization run, and the final minimum $\chi^2$ is chosen as the minimum of the minima of each chain, making the result more stable. 
We explore the differences between the two optimizers Minuit and our SA algorithm in Sections \ref{subsec:min_validation} and \ref{subsec:compare_mcmc_minuit}.

\subsection{Initialization}
\label{subsec:initialization}

Most optimization methods, including SA, require an initial point to begin from. 
Moreover, as Procoli piggy-backs over an MCMC to explore the parameter space, converging to a maximum $\mathcal{L}$ (or minimum $\chi^2$) is substantially faster and more accurate with an input covariance matrix $\Sigma$. 
Both of these requirements are easily fulfilled by initializing the profile likelihood run with a converged MCMC. 
This provides a maximum a posteriori (MAP) in the form of the global maximum likelihood found by the MCMC, as well as a covariance matrix for the paraemter space. 
The convergence criteria for these input MCMC chains is not very strict, a Gelman-Rubin critera of $R-1 < 0.2$ suffices. 
Alternatively, supplying just an input covariance matrix and MAP also fulfills these requirements. 

Using these, the code first finds the global best fit by starting from the MAP and using the covariance matrix, carrying out an SA optimization as described above. 
The recommendations for $T$ and $\delta$ for the global optimizer are more thorough, as parameter confidence intervals from profile likelihoods are found at $\Delta \chi^2 = 1$ relative to the global minimum $\chi^2_{\min}$. 

Once this global minimum is determined, the code takes iterative steps away from this minimum along either direction in the profile likelihood parameter $\theta_i$, at a user-defined step-size $\Delta \theta_i$. 
That is, after the global best fit $\bm{\theta}_{\min} = \bm{\theta}^0$ is found, the code fixes 
\begin{equation}
    \theta_i = \theta_i^{+1} \equiv \theta_{i, \min} + \Delta \theta_i \,. 
\end{equation}
The superscripts denote the iteration, with $0$ corresponding to the global minimum $\chi^2$. 
It then initializes the SA optimizer providing $\bm{\theta}^0$ as initial value for all other parameters $\theta_{j \neq i}$, and the covariance matrix $\Sigma$ from the MCMC run. 
The optimizer finds a new minimum $\bm{\theta}^{+1}$, which becomes the input for the next iteration. 
This sequence continues, with each iteration fixing 
\begin{align}
    \theta_i = \theta_i^{+n} &\equiv \theta_i^{0} + n\Delta \theta_i \nonumber \\
    \theta_{j \neq i} &\in \bm{\theta}^{+(n-1)} \,.
\end{align}
A similar sequence progresses in the negative direction for $\theta_i^{-n}$ and $\bm{\theta}^{-n}$, continuing till the user-defined ranges for exploration in $\theta_i$ are saturated, resulting in a sequence of optimized points $\{\bm{\theta}^{-N}, ... \bm{\theta}^{-1}, \bm{\theta}^0, \bm{\theta}^{+1}, ... \bm{\theta}^{+M} \}$. 
Note that $N \neq M$ usually, as the global best fit $\bm{\theta}^0$ may be closer to one bound. 

For this sequential profile likelihood of $\theta_i$, we recommend that the first jumping factor  $\delta$ in the SA optimizer be $\delta \leq \Delta \theta_i / \sigma_i$, where $\sigma_i$ is the $1\sigma$ error on the likelihood-profile parameter $\theta_i$ from the covariance matrix. 
This avoids scenarios where the optimizer might overshoot the next minimum, and instead allows the optimizer to converge to this minimum. 

This procedure records not only the profile likelihood, but also the values of all input and derived parameters across the profile, including a breakdown of $\chi^2$ per experiment.

\subsection{Implementation}
\label{subsec:implementation}

The optimization method and sequential calculation of the likelihood profile described above are implemented by Procoli in Python. 
The Procoli Python package relies primarily on MontePython to perform the SA optimization via running MCMCs through subprocesses, and to analyze the chains produced to find the best fit and calculate the covariance matrix. 
It also uses functions defined in GetDist \cite{Lewis:2019xzd} to check any provided MCMC chains and their convergence, but this is not necessary as the user can alternatively provide just a covariance matrix and a starting guess for the global best fit. 
The code then takes as input: 
\begin{itemize}
    \item \verb|chains_dir|: path to the directory for the location of the MCMC chain files or alternatively, the covariance matrix (\verb|.covmat|) and best-fit guess (\verb|.bestfit|) files, in the format used by MontePython. 
    This directory must also contain a \verb|log.param| file that includes the experiments, and cosmological and nuisance parameters to be used, identical to a usual MontePython MCMC run. 
    \item \verb|prof_param|: name of the profile likelihood parameter $\theta_i$ as understood by MontePython, same as in the \verb|log.param|
    \item \verb|prof_incr| increment $\Delta \theta_i$ at which to query the profile
    \item \verb|prof_min| and \verb|prof_max|: the range in $\theta_i$ to explore 
\end{itemize}
Additionally, the user can also specify the number of chains to run in parallel for each optimization by setting \verb|processes| (default assumed to be 5). 
This is effectively the number of MCMC chains used in the simulated annealing. 
A complete list of arguments is documented in the code on GitHub. 

The code begins by looking for MCMC chain files or for the covariance matrix and input guess for the best fit. 
If only the chains are provided, it obtains the covariance matrix and best-fit guess using MontePython analysis tools. 
It then begins a global optimizer using any specified lists of values for the jumping factor and temperature or uses the defaults. 
Next, it begins the profile likelihood exploration using the input increments $\Delta \theta_i = $ \verb|prof_incr|, till it saturates either \verb|prof_min| or \verb|prof_max|. 
The jumping factor and temperature lists for the profile differ from those for the global best fit search, and both can be modified by the user. 

At each iteration, the code populates a sub-directory within the parent directory with the MCMC chains run by the optimizer. 
These chains are overwritten per iteration $\Delta \theta_i$ to save memory space. 
A text file in the parent directory keeps track of the likelihood profile and the location of the run. 
If prematurely terminated, the code restarts the profile at the next point after the last complete iteration. 
Users can keep track of how long each iteration takes, and therefore how long the entire run would take via handy time-stamp files produced. 

Examples for a complete session are recorded on the GitHub, using Jupyter notebooks as well as a python script that can be submitted to a cluster. 
Procoli also includes several useful plotting and analysis functions, which are illustrated in these Jupyter notebooks.

\section{Examples and features} 
\label{sec:examples}

For the purposes of demonstrating the code, we explore two cosmological models, \LCDM\ and early dark energy (EDE) \cite{Karwal:2016vyq,Poulin:2018cxd,Poulin:2018dzj,Poulin:2023lkg,Kamionkowski:2022pkx}. 
EDE adds an additional component to base \LCDM\ that behaves like a cosmological constant at early times, then close to matter-radiation equality, rapidly dilutes away, such that its impact on cosmology is localized in redshift. 
It was originally proposed as a solution to the Hubble tension \cite{Verde:2019ivm,Abdalla:2022yfr,Kamionkowski:2022pkx}, as it shifts the CMB-predicted $H_0$ into agreement with local distance-ladder measurements \cite{Riess:2023egm, Riess:2022mme}. 
The CMB predicts $H_0$ via a very precise measurement of the angular size $\theta_*$ of the sound horizon $r_s$ at the surface of last scatter \cite{Karwal:2016vyq,Bernal:2016gxb,Aylor:2018drw,Knox:2019rjx,Poulin:2023lkg,Kamionkowski:2022pkx}. 
Approximately, $\theta_* \propto H_0 r_s$. 
The addition of EDE makes the pre-recombination universe expand slightly faster, shrinking $r_s$, which increases $H_0$ to fix the observed value of $\theta_*$. 

\begin{figure}
    \centering
    \includegraphics[width=0.4\textwidth]{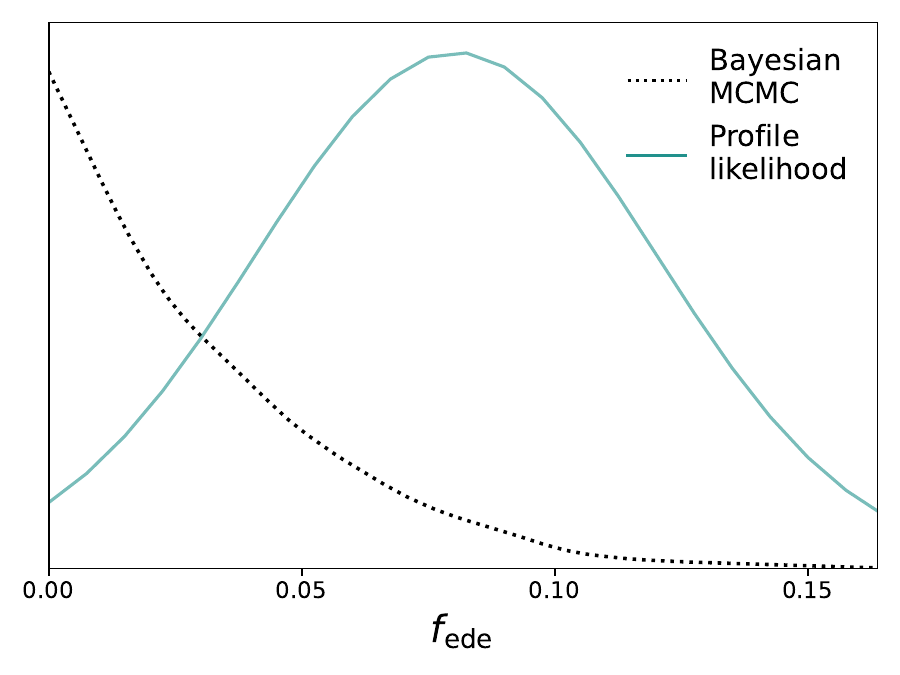}
    \caption{
    Bayesian posteriors on the amount $\fede$ of EDE are shown (dotted black) alongside a likelihood profile (solid green) on the same parameter, using the same data sets, namely Planck CMB primaries. 
    EDE posteriors from Bayesian MCMCs are impacted by prior volume growing in the direction of $\fede \rightarrow 0$, biasing constraints towards $\fede = 0$, despite the maximum likelihood being closer to $\fede \simeq 0.07$ as seen from the maximum of the solid blue curve. 
    }
    \label{fig:ede_prior_vol_effects}
\end{figure}

Numerous models can produce the EDE phenomenology \cite{Poulin:2018dzj, Lin:2019qug, Agrawal:2019lmo, Niedermann:2019olb, Karwal:2021vpk, Berghaus:2019cls, Berghaus:2022cwf, McDonough:2021pdg, Brissenden:2023yko, Braglia:2020auw, Gonzalez:2020fdy, Sakstein:2019fmf}. 
Here we concentrate on ultra-light-axions-inspired EDE, using AxiCLASS\footnote{
\href{https://github.com/PoulinV/AxiCLASS/tree/master}{AxiCLASS} is publicly available on GitHub
} \cite{Poulin:2018cxd,Poulin:2018dzj}. 
The parameters of this model are the maximum fractional energy density $\fede$ in EDE, 
the redshift $z_c$ (or alternatively the scale factor $a_c$) at which EDE density peaks, 
and the initial value $\phi_i$ of the scalar field that forms EDE\footnote{
Usually, this parameter is denoted by $\theta_i$, but we use $\phi_i$ here to distinguish it from the discussion in Sec.~\ref{sec:code}}. 
As $\fede \rightarrow 0$, the properties of this negligible component defined by $z_c$ and $\phi_i$ become unconstrained. 
Then, the choice of uniform priors in $\log_{10}z_c$ and $\phi_i$ become informative priors, rapidly growing the prior volume in the direction $\fede \rightarrow 0$ and biasing Bayesian posteriors as shown in Fig. \ref{fig:ede_prior_vol_effects}.

Besides the EDE parameters, we also vary the six \LCDM\ parameters: 
the physical densities of cold dark matter $\omega_c$ and baryons $\omega_b$, 
the amplitude $A_s$ and tilt $n_s$ of the primordial power spectrum, 
the optical depth $\tau_{\rm reio}$ to reionization and the Hubble expansion rate $H_0$ today. 

Our primary example fits an EDE cosmology to the Planck 2018 CMB spectrum including the low-$\ell$ TT and EE data and the high-$\ell$ TTTEEE data \cite{Planck:2019nip}. 
Some of our demonstrations additionally include the Planck CMB lensing spectrum \cite{Planck:2018lbu}, supernova data from Pantheon \cite{Pan-STARRS1:2017jku}, as well as BAO data from BOSS DR12 \cite{BOSS:2016wmc}, 6dF \cite{Beutler:2011hx} and SDSS MGS \cite{Ross:2014qpa}. 
Beyond these, any likelihoods available in MontePython can be readily used in Procoli. 

With this setup, we perform several tests to check the robustness of our code, and to provide recommendations for usage. 
Specifically, in the following, we demonstrate the accuracy of the global optimizer with an example using mock data generated for an EDE cosmology in \ref{subsec:min_validation}. 
We next compare our optimizer to simply binning the MCMC chains and looking for the MAP within bins, and to using Minuit, a deterministic optimizer for the profile in \ref{subsec:compare_mcmc_minuit}. 
Besides these tests, we also showcase a key feature of Procoli  in \ref{subsec:chi2_per_exp} - extracting the $\chi^2$ per experiment, which can provide deeper insights into data with profile likelihoods.

\subsection{Accuracy of the global best fit}
\label{subsec:min_validation}

\begin{table*}[]
    \centering
    \begin{tabular}{l|c|c|c|c|c|c|c|c|c}
    Parameter                
    &  \shortstack{Fiducial \\ value} 
    & \shortstack{Input \\ MAP} & $\Delta \theta^{\rm input}_i / \theta_i^{\rm fid}$  
    & \shortstack{SA \\ best fit} & $\Delta \theta^{\rm SA}_i / \theta_i^{\rm fid}$  
    & \shortstack{Input \\ MAP} & $\Delta \theta^{\rm input}_i / \theta_i^{\rm fid}$  
    & \shortstack{SA \\ best fit} & $\Delta \theta^{\rm SA}_i / \theta_i^{\rm fid}$  \\    
    \hline
    \hline
$\omega_{b} $	 & 	 0.02291 	 & 	 0.02262 	 & 	 -0.01 & 	 0.02291 	 & 	 0.0001 & 	 0.02275 	 & 	 -0.007 & 	 0.02291 	 & 	 -0.0001 \\
$\omega_{cdm} $	 & 	 0.1278 	 & 	 0.12724 	 & 	 -0.004 & 	 0.12785 	 & 	 0.0004 & 	 0.13199 	 & 	 0.03 & 	 0.12777 	 & 	 -0.0003 \\
$h$	 & 	 0.7051 	 & 	 0.69737 	 & 	 -0.01 & 	 0.70532 	 & 	 0.0003 & 	 0.71967 	 & 	 0.02 & 	 0.70494 	 & 	 -0.0002 \\
$\ln 10^{10}A_{s} $	 & 	 3.049 	 & 	 3.06196 	 & 	 0.004 & 	 3.04903 	 & 	 1e-05 & 	 3.07104 	 & 	 0.007 & 	 3.04879 	 & 	 -7e-05 \\
$n_{s} $	 & 	 0.9771 	 & 	 0.98001 	 & 	 0.003 & 	 0.97726 	 & 	 0.0002 & 	 0.99013 	 & 	 0.01 & 	 0.97696 	 & 	 -0.0001 \\
$\tau_{\rm reio } $	 & 	 0.0593 	 & 	 0.05674 	 & 	 -0.04 & 	 0.05929 	 & 	 -9e-05 & 	 0.05714 	 & 	 -0.04 & 	 0.05924 	 & 	 -0.001 \\
$f_{\rm ede} $	 & 	 0.1173 	 & 	 0.07319 	 & 	 -0.4 & 	 0.11781 	 & 	 0.004 & 	 0.12271 	 & 	 0.05 & 	 0.11691 	 & 	 -0.003 \\
$\log_{10} a_c $	 & 	 -3.614 	 & 	 -3.57397 	 & 	 -0.01 & 	 -3.61403 	 & 	 9e-06 & 	 -3.56097 	 & 	 -0.01 & 	 -3.61293 	 & 	 -0.0003 \\
$\phi_{i}$	 & 	 2.689 	 & 	 2.72609 	 & 	 0.01 & 	 2.68889 	 & 	 -4e-05 & 	 2.76404 	 & 	 0.03 & 	 2.68805 	 & 	 -0.0004 \\
$\chi^2$	 & 	                 & 	 591.83 	 & 	                 & 	 0.00054 	 & 	                 & 	 1162.78 	 & 	                 & 	 0.00087 	 & 	                 
    \end{tabular}
    \caption{
    The parameters of the fiducial cosmology used to generate the mock data are compared to the parameters found in the global best fit using the Procoli default SA optimizer. 
    We also show relative differences $(\theta_i - \theta_i^{\rm fid})/\theta_i^{\rm fid}$ for each column. 
    SA best fits from two different starting points are shown - the first input MAP comes from an EDE fit to Planck data, with its distance from the true best-fit fiducial value captured by its $\chi^2$ for the mock data in the last row. 
    The second input MAP comes from an EDE fit to Planck CMB + Pantheon supernovae + BAO + SH0ES data, a substantially worse starting guess as seen from its $\chi^2$ for the mock data. 
    Each of these input MAPs are followed by their best fits found through the SA optimizer. 
    }
    \label{tab:ede_validation}
\end{table*}

To validate the global optimizer, we generated noiseless mock CMB data up to multipole $\ell = 3000$ in a fiducial EDE cosmology using the parameters shown in Table~\ref{tab:ede_validation}. 
Then, using the default settings for the SA optimizer in Procoli, we find a global best fit for this mock data. 
Two different inputs are tested for the initial best-fit guess and covariance matrix. 
One from an EDE fit to just Planck data, with an initial guess that has $\chi^2 = 592$ for the mock data. 
The second input is taken from an EDE fit to Planck CMB, Pantheon supernovae, BAO and SH0ES data, with a comparatively worse $\chi^2=1163$ for the mock data. 
Although these are not optimal inputs as they are based on different data, this setup provides a useful test case to validate the optimizer searching for the global best fit. 
Doing so, we nearly exactly recover the true cosmology and find $\chi^2_{\min} = 0.00054$ in the case of the smaller input $\chi^2$ and $\chi^2_{\min} = 0.00087$ for the second case, with differences of $\Delta \chi^2 < 0.001$ between the two scenarios. 
Hence, even with sub-optimal inputs, the SA optimizer is excellent at recovering the true parameter values. 

These scenarios effectively stress-test the SA optimizer. 
Usually, global optimizers are initialized from an MCMC run on the same data sets, providing better input MAPs and covariance matrices, scenarios which are only expected to improve on the performance above. 
For the sake of comparison, a quasi-Newtonian optimizer subjected to the same atypical scenario above could not recover the true inputs. 
We test Minuit under the same conditions, with the same sub-optimal guesses for the initial point and covariance matrix and it recovers $\chi^2_{\rm min} = 255.86$ and $\chi^2_{\rm min} = 266.71$ respectively in the two cases. 
Here, we use the \verb|iminuit.minimize| function as a swap for \verb|scipy.optimize.minimize| in MontePython, as these have the same interface, although this does not allow utilization of the full functionality of Minuit. 
Also note that this is not usually how Minuit is employed to optimize - relying on an input covariance matrix and starting guess for the best fit that were not generated for the data set in question. 
We simply highlight that Minuit is more sensitive to the initial guess and these unusual conditions form a critical impediment to it, unlike SA whose stochasticity can overcome a bad initial starting point. 
Beyond profile likelihoods, we recommend the use of SA optimizers for maximum-likelihood searches with real CMB data, which is far noisier and higher-dimensional than this test case. 
The stochasticity of SA optimizers leads to more robust best fits than deterministic algorithms like quasi-Newton methods. 

\begin{figure}
    \centering
    \includegraphics[width=0.49\textwidth]{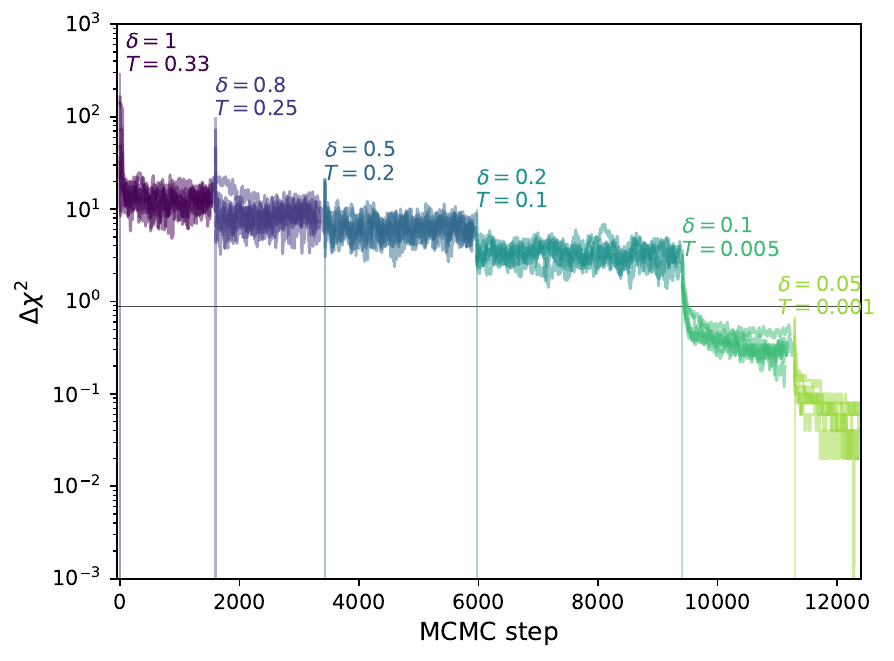}
    \caption{
    The $\Delta \chi^2$ relative to the minimum $\chi^2_{\rm min} = 2761.3$ for a global optimizer run using the default settings in Procoli for the jumping factor $\delta$ and temperature $T$ for the SA ladder. 
    We separate the steps by color and vertical lines and denote the values of $\delta$ and $T$ used. 
    Each chain at each step in the ladder took a total of 5000 MCMC steps, including the rejected steps. 
    The horizontal line marks the minimum global $\chi^2$ achieved by Minuit, $\Delta \chi^2_{\rm min} = 0.89$ greater than the Procoli minimum. 
    The run was for an EDE cosmology constrained by Planck CMB primaries. 
    }
    \label{fig:global_min}
\end{figure}

In Fig.~\ref{fig:global_min}, we show a global optimizer run for an EDE cosmology with real data, namely Planck CMB primaries. 
For this, we use the default settings in Procoli for the jumping factor and temperature progression over the SA optimization, as labeled on the figure. 
By default, each chain for each step in the SA ladder takes $N=4000$ MCMC steps, including rejected steps, but here we run the optimizer slightly longer with $N=5000$. 
For each SA step, multiple chains are run in parallel to explore the parameter space, further capitalizing on the stochastic nature of SA.
Then, the best-fit point of all previous SA steps across all chains is used as input for the next SA step. 
As each step has a finite jumping factor $\delta$, the optimizer can still escape local minima. 
As the temperatures $T$ and jumping factors are slowly lowered, the differences in the $\chi^2$ of chains follow suit. 
Ultimately, the optimizer finds the maximum global likelihood with multiple chains. 
EDE and \LCDM\ parameter values per MCMC step are shown in Appendix \ref{sec:appendix_plots}.

We again compare the results for a global best fit found using Minuit to our SA optimizer, and find that Minuit results in only a slightly worse fit by $\Delta \chi^2_{\rm min} = 0.89$. 
Minuit performs far better in this case when its inputs are generated for the data set in question, the usual scenario in which Minuit is run post an MCMC. 
However, for profile likelihoods that set the $1\sigma$ parameter limits at $\Delta \chi^2 = 1$, this difference in $\chi^2$ is substantial, making SA optimizers the superior choice.

\subsection{Comparison with Minuit and binned MCMC }
\label{subsec:compare_mcmc_minuit}

\begin{figure}
    \centering
    \includegraphics[width=0.49\textwidth]{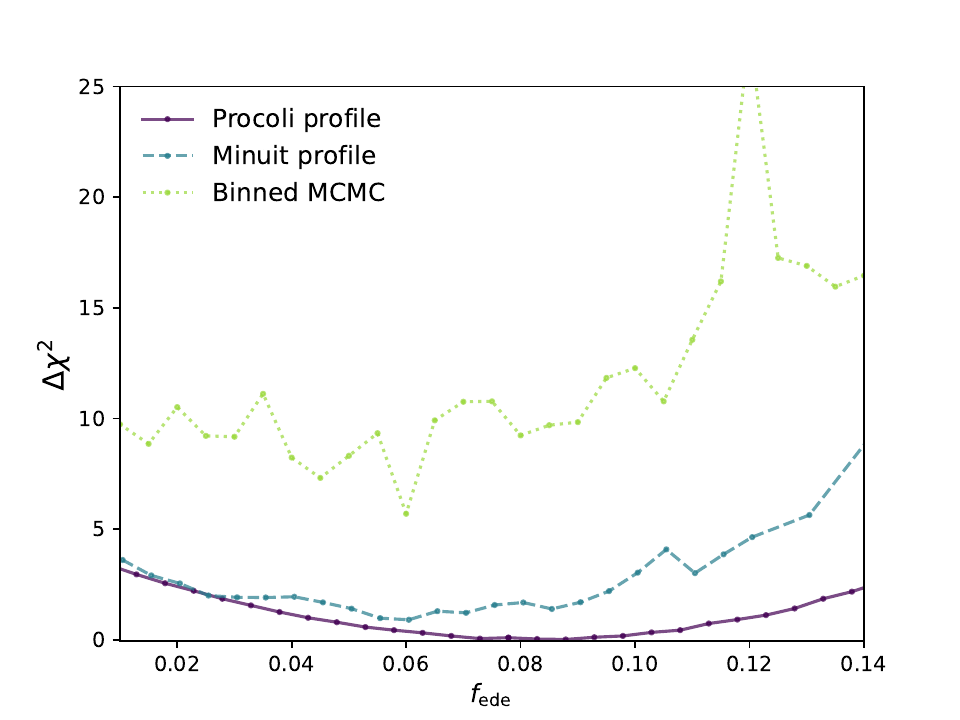}
    \caption{
    Profile likelihoods of Planck CMB primaries for $\fede$ obtained via three approaches are compared. 
    For solid purple, we use Procoli to produce a profile likelihood, using its SA optimizer. 
    For dashed blue, we use the same sequential approach described in Section \ref{subsec:initialization}, but use Minuit instead of SA. 
    In dotted green, we bin the samples from a vanilla Metropolis-Hastings MCMC in narrow bins centered at a given $\fede$, picking the MAP per bin. 
    To rescale them and aid visuals, we subtract out the minimum $\chi^2=2761.21$ found across all three profile from each curve. 
    }
    \label{fig:mcmc_min_comparison}
\end{figure}

Beyond just the global best fits, the SA and quasi-Newtonian optimizers can also be compared on the full profile likelihoods, as we do in Fig.~\ref{fig:mcmc_min_comparison}. 
Alongside, we also compare simply binning the points sampled by an MCMC in narrow bins centered at the same $\Delta \theta_i$ increments as the profile likelihoods, picking the MAP per bin. 

\begin{figure}
    \centering
    \includegraphics[width=0.4\textwidth]{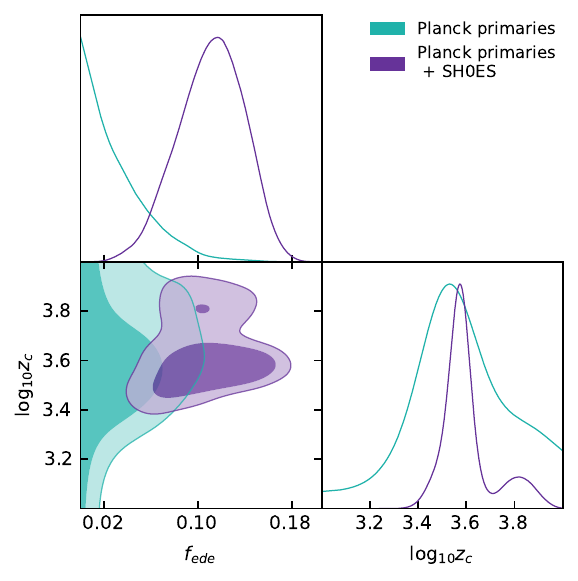}
    \caption{Any bimodality in $z_c$ is very mild when EDE is constrained with just Planck CMB primaries, indeed the two modes are connected within the 1$\sigma$ posterior contour as shown by the green contours. 
    The bimodality is sharpened and more clearly visible with the inclusion of the local $H_0$ measurement \cite{Riess:2021jrx} in the purple contours.
    }
    \label{fig:zc_bimod_mcmc}
\end{figure}

The profile based on a Metropolis-Hastings MCMC is mildly biased towards lower $\fede$, partly due to being afflicted by prior volume effects. 
Moreover, the profile is extremely jagged and the MAPs in each bin are $>4$ in $\chi^2$ away from the minima found by Procoli. 
Basing a profile likelihood on simply binning MCMC posteriors can not only be misleading, but may also completely miss the true minimum, as is the case for the global MAP of the full MCMC chains. 
In particular, the global MAP finds $\log_{10}z_c = 3.82$, while the global best fit found by Procoli as described in Sec. \ref{subsec:min_validation} finds $\log_{10}z_c = 3.57$. 
This arises from a mild bimodality in EDE cosmologies in the injection redshift $z_c$ of EDE, shown in Fig. \ref{fig:zc_bimod_mcmc}, which is recovered by a standard MH MCMC that converges within finitie time. 
The bimodality also forms an additional factor biasing the binned MCMC curve in Fig. \ref{fig:mcmc_min_comparison} towards lower $\fede$. 

\begin{figure}
    \centering
    \includegraphics[width=0.49\textwidth]{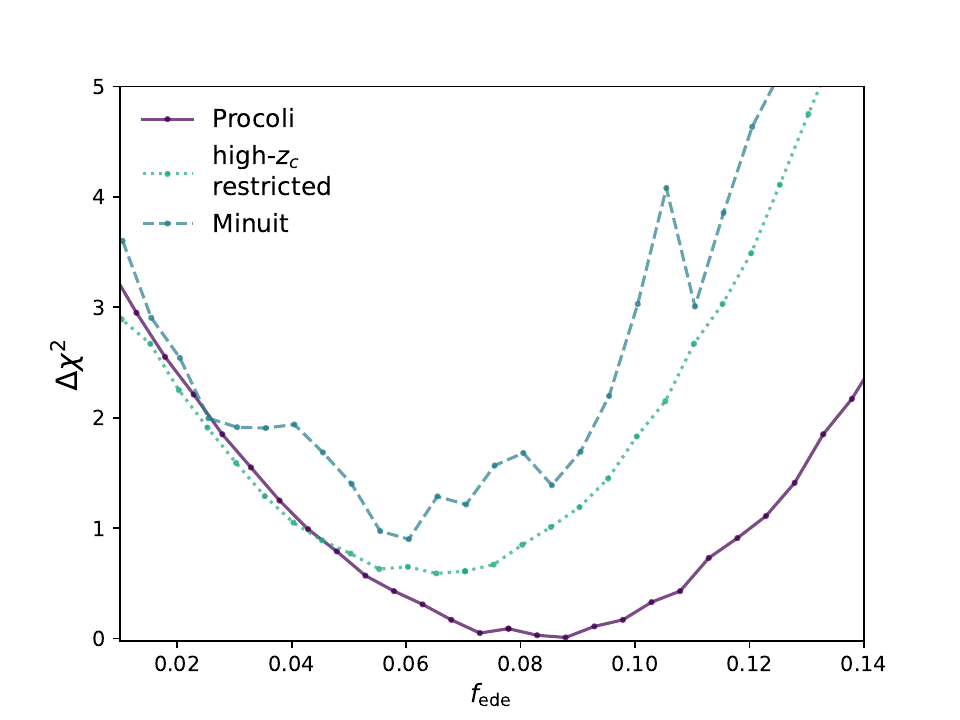}
    \caption{
    We compare the profile likelihoods of Planck CMB primaries for $\fede$ obtained using two different optimizers - SA via Procoli and a quasi-Newton method via Minuit. 
    Minuit (dashed blue) gets stuck in a local minimum with EDE injection redshift $\log_{10}z_c > 3.7$, which is not the true global minimum. 
    Procoli (solid purple) finds a better minimum at $\log_{10}z_c < 3.7$, 
    but a profile resembling the Minuit profile can be recovered by Procoli by restricting it to this high-$z_c$ mode (dotten green).
    The solid purple Procoli curve and the dashed blue Minuit curves are the same as in Fig. \ref{fig:mcmc_min_comparison}. 
    }
    \label{fig:compare_minimizers}
\end{figure}

Using the deterministic optimizer Minuit, we also perform the same sequential search for the likelihood profile as described in Sec. \ref{subsec:initialization}, but swapping the SA optimizer for Minuit. 
The same chains as the binned MCMC curve are used for input for this curve, i.e. Minuit is initialized at a MAP that tracks the higher-$z_c$ mode, instead of the lower-$z_c$ mode which covers the true global best fit. 
Although this setup performs far better than a binned MCMC, it does not escape the high-$z_c$ local minimum which it tracks for all $\fede$, failing to find the true best fit.
A similar profile is recovered by Procoli if $z_c$ is restricted to this mode with $\log_{10}z_c > 3.7$ as shown in Fig. \ref{fig:compare_minimizers}. 

Finally, with the same inputs as the binned MCMC and the Minuit curves, the SA optimiser in Procoli finds the true best fit in the lower-$z_c$ mode. 
Its stochastic approach is able to overcome the mild bimodality of EDE. 
Moreover, it finds a smoother profile with consistently lower $\chi^2$ values. 

This example with EDE does not generalize divergent results from quasi-Newtonian and SA optimizers. 
It simply highlights the possible pitfalls of deterministic optimizers when employed over a complex and high-dimensional parameter space and urges caution when using them to produce likelihood profiles. 

\subsection{Differentiating $\chi^2$ per experiment}
\label{subsec:chi2_per_exp}

\begin{figure*}
    \centering
    \includegraphics[width=0.49\textwidth]{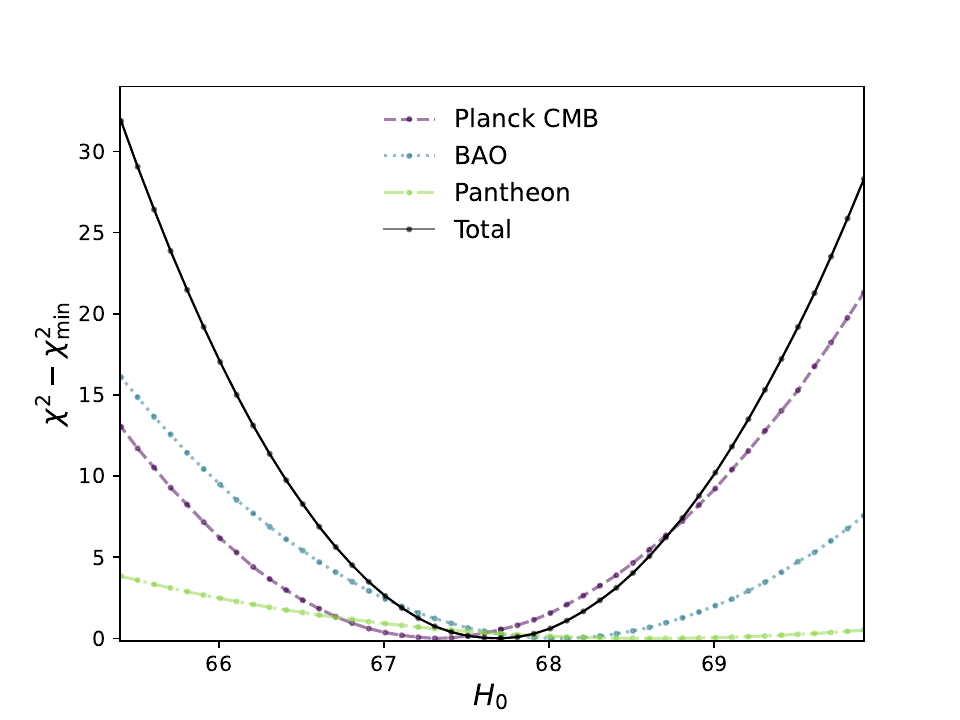}
    \includegraphics[width=0.49\textwidth]{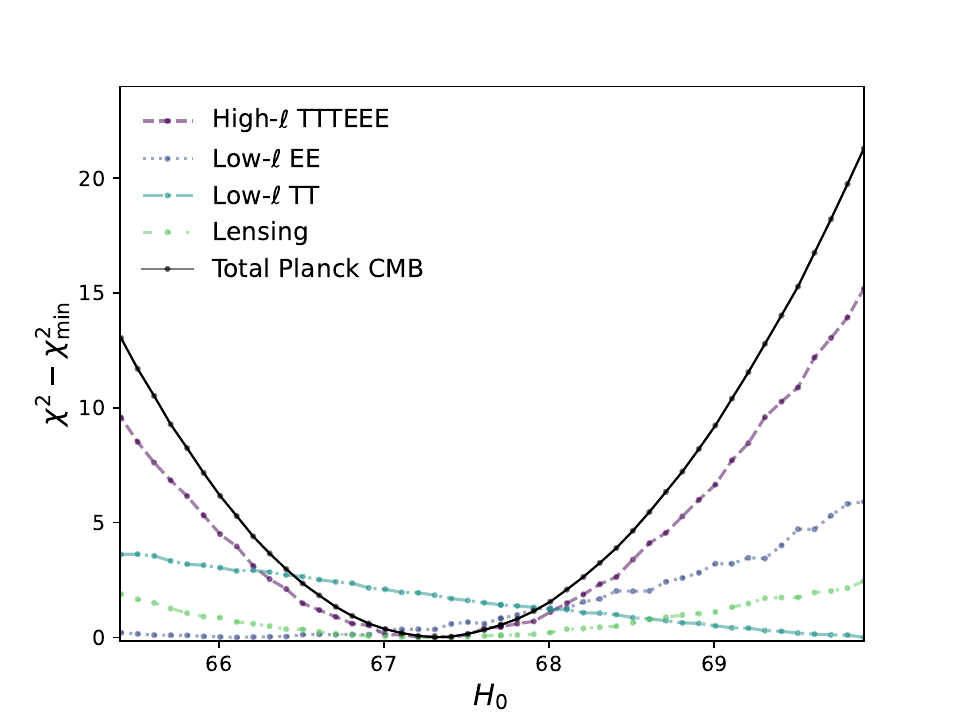}
    \caption{
    A breakdown of the $\chi^2$ per experiment for a profile likelihood on $H_0$ in \LCDM, fitting to the experiments listed. 
    For each curve, we subtract out its minimum $\chi^2$ to show them on the same scale. 
    \textit{Left:} The full Planck CMB dataset is shown in dashed purple, BAO in dotted blue and Pantheon supernovae in dash-dotted green. 
    The total $\chi^2$ shown in solid black is the quantity optimized for producing this profile. 
    \textit{Right:} We further breakdown the elements contributing to the Planck CMB curve on the left plot (same as solid black here), 
    namely the high-$\ell$ TTTEEE spectrum (dashed purple), 
    the low-$\ell$ EE (dotted blue) and TT (dash-dotted teal), 
    and the CMB lensing power spectrum (dash-dot-dotted green). 
    }
    \label{fig:chi2_per_exp}
\end{figure*}

For each point in the likelihood profiles produced by Procoli, the code also calculates the $\chi^2$ for each experiment listed in the \verb|log.param|. 
This can provide powerful insights into which experiments are driving constraints and also automatically allows the user to produce profiles using a subset of experiments as shown in Fig. \ref{fig:chi2_per_exp}. 

For demonstrating this, we profile $H_0$ in a \LCDM\ model constrained by Planck CMB, BAO and Pantheon supernova data. 
The left panel of Fig. \ref{fig:chi2_per_exp} shows the profiles for these three data sets and their cumulative total $\chi^2$, with each curve rescaled by subtracting out the minimum $\chi^2$ for that specific curve. 
This $\chi^2 - \chi^2_{\rm min}$ exactly quantifies the contribution of each data set to the combined profile likelihood, as shown by the solid black curve balancing the individual profiles of BAO and Planck CMB data. 
It is also immediately clear which datasets are most constraining for the profile parameter (Planck CMB), and the one with minimal impact to the profile (Pantheon supernovae). 
While this is no surprise for the thoroughly studied \LCDM\ model, such splits differentiating the $\chi^2$ each experiment contributes to likelihood profiles can provide valuable insights into more complicated models. 

For the left panel in Fig. \ref{fig:chi2_per_exp}, we combined individual CMB likelihoods under Planck CMB, but these can also be examined separately as in the right panel. 
The solid black curve in the right panel now corresponds to the combined Planck CMB likelihood, and is the same as the dashed purple curve in the left plot. 
For just Planck CMB data, these profiles result in a constraint $H_0 = 67.3 \pm 0.54$ km/s/Mpc, while the combined data sets from the left panel result in the slightly more stringent constraint $H_0 = 67.7 \pm 0.42$ km/s/Mpc, as expected with the addition of more data. 
Simply by considering just a subset of the experiments from the left panel, one can automatically produce a profile on $H_0$ with only Planck CMB data, effectively getting multiple profiles for the price of one. 

Another notable feature is the varying level of smoothness of each curve particularly in the right panel of Fig. \ref{fig:chi2_per_exp}. 
For example, from this one can read off that the wrinkles in the low-$\ell$ EE data contribute the most noise to the Planck CMB profile, and any optimizations over Planck CMB data. 
Such $\chi^2$ splits over experiments hence also provide insight into the data itself.

\section{Conclusions}
\label{sec:conclusions}

Bayesian and frequentist statistics are two alternative approaches to parameter estimation with data. 
Dedicated work by cosmologists has built up a strong arsenal of Bayesian tools, powerful at incorporating prior information on model parameters. 
The explicit reliance on priors can however leave Bayesian approaches susceptible to prior-volume effects that bias results, including in early dark energy cosmologies, the example used in this work. 
As we search for new physics beyond \LCDM\ and for the true fundamental model of the Universe, it is imperative that our parameter estimations be free of bias. 

Frequentist statistical tools such as profile likelihoods form a means of obtaining parameter estimates free of priors, providing a valuable supplement to Bayesian constraints. 
These contribute additional insight into both the data and theoretical models, with constraints that have fundamentally complementary interpretations. 
Bayesian constraints are statements about the posterior, while frequentist constraints constitute statements about the data likelihood. 
With cosmic anomalies hinting at new physics and an immense influx of high-resolution cosmological data expected in the coming years, it is crucial to build up the framework for alternative statistical approaches in cosmology. 

This work presents Procoli, a straightforward and fast publicly available Python package to obtain profile likelihoods in cosmology. 
Procoli wraps MontePython, a sampler written for use with the CLASS Boltzmann code, automatically incorporating not only all data likelihoods available for use with MontePython, but also all modified versions of CLASS that implement physics beyond the standard \LCDM\ model. 
Here we describe the code and demonstrate its features. 
Procoli calculates profile likelihoods sequentially, maximizing the likelihood at each point over the profile parameter, using the previous adjacent point to initialize the next. 
At their core, profile likelihoods depend on reliable optimizers. 
We validate the simulated-annealing optimizer used in Procoli with mock EDE data, and find it to be more robust and stable that a quasi-Newtonian method on comparison with Minuit, both for finding a global best fit and for the profile itself. 
Finally, we demonstrate Procoli functionality for dissecting the $\chi^2$ per experiment to gain additional insight into the data and the cosmological model. 
Besides these illustrations of Procoli, we also provide example running scripts, output of completed runs and basic plotting functionality on the GitHub. 

During the final development of this code, the PROSPECT code \cite{Holm:2023uwa} was released, another tool to extract profile likelihoods in cosmology. 
Both Procoli and PROSPECT use simulated-annealing optimizers, but these are implemented differently. 
Procoli sequentially constructs the likelihood profile and currently has a set simulated-annealing ladder of temperature and step-size progressions, to be updated to an adaptive ladder in the next version. 
Each point in the profile is initialized from a previously-optimized, adjacent point such that inputs to the SA optimizer are close to the maximum likelihood already. 
PROSPECT approaches the profile in parallel and implements an adaptive simulated-annealing algorithm. 
As input for each point in the profile, it takes MAPs from the MCMC chains in narrow bins, similar to those shown in Fig. \ref{fig:mcmc_min_comparison}. 
With frequentist profile likelihoods regaining popularity in cosmology, having multiple codes to perform similar tasks can be extremely useful both for comparison and accessibility. 

Bayesian and frequentist statistical approaches each have strong advantages. 
Employing them in conjunction can yield deeper, unbiased insights into data as well as cosmological models and lead to a better understanding of the impact of prior choices. 
Procoli contributes to this goal, adding to the growing arsenal of frequentist tools at the disposal of cosmologists.

\acknowledgements{
We thank Marco Raveri, Cyrille Doux, Shivam Pandey, Jos\'e Luis Bernal and Nils Sch\"oneberg for discussions that helped improve this work. 
We additionally thank Shar Daniels for their input during the early stages of developing this code. 
Finally, we are grateful to Marco Gatti for contributing computing resources for the completion of this project. 
TK was supported by funds provided by the Kavli Institute for Cosmological Physics at the University of Chicago through an endowment from the Kavli Foundation. 
VP is supported by funding from the European Research Council (ERC) under the European Union’s HORIZON-ERC-2022 (grant agreement no.~101076865). 
VP is also supported by the European Union’s Horizon 2020 research and innovation program under the Marie Sk{\l}odowska-Curie grant agreement no.~860881-HIDDeN. 
TLS is supported by NSF Grant No.~2009377. 
The authors acknowledge the use of computational resources from the Excellence Initiative of Aix-Marseille University (A*MIDEX) of the Investissements d’Avenir program and from the LUPM's cloud computing infrastructure founded by Ocevu labex and France-Grilles.
}

\bigskip

\appendix

\section{Global optimization convergence}
\label{sec:appendix_plots}

\begin{figure*}
    \centering
    \includegraphics[width=0.45\textwidth]{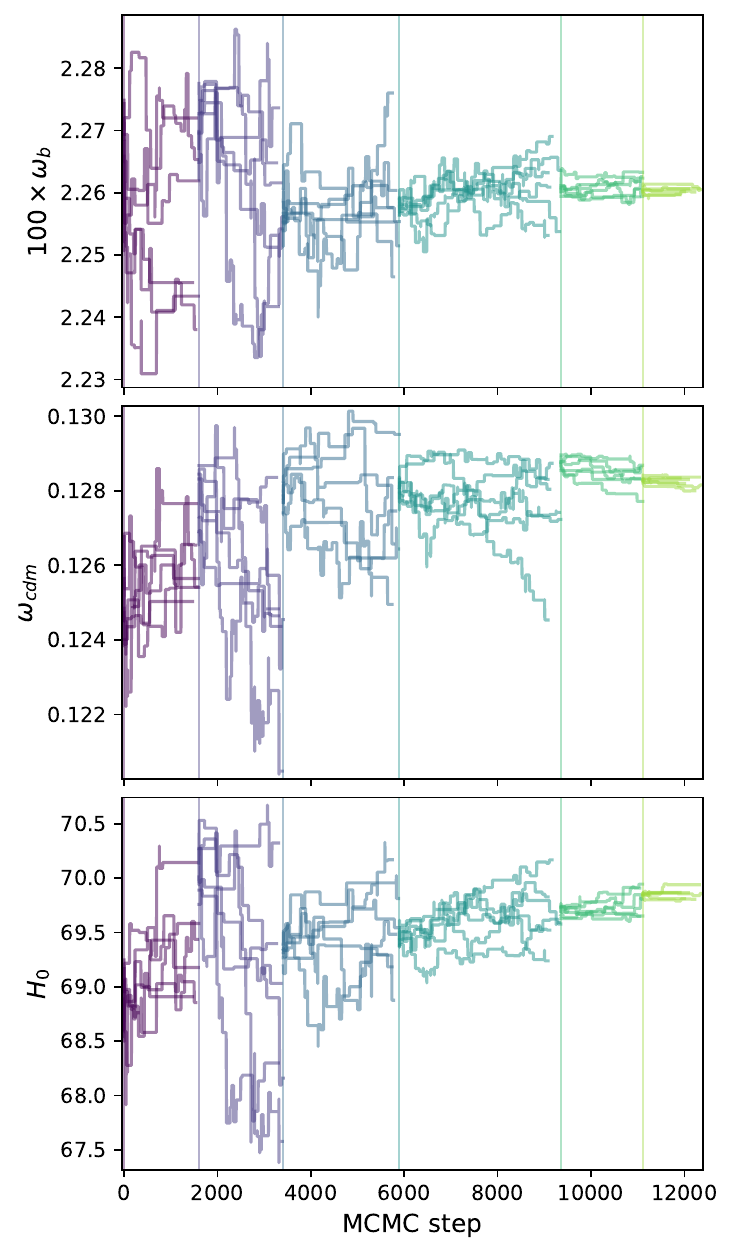}
    \includegraphics[width=0.458\textwidth]{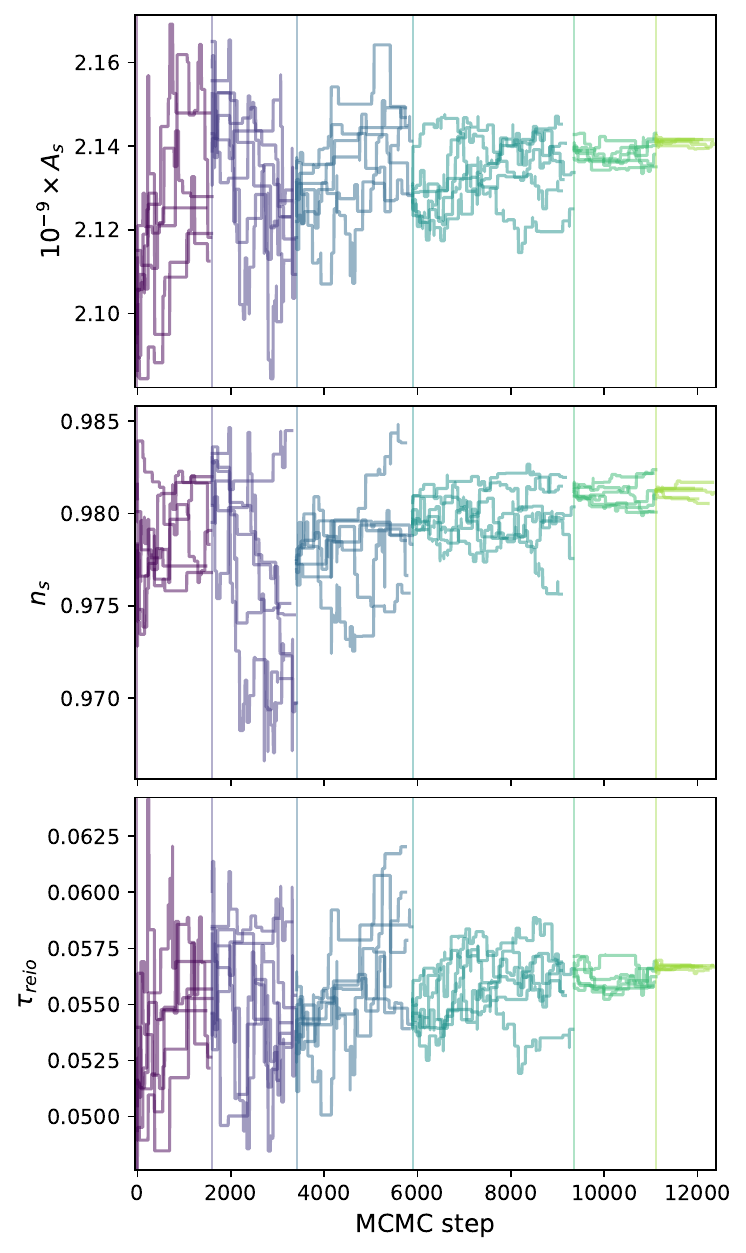}
    \caption{
    Convergence of the 6 \LCDM\ parameters over the global optimizer run in an EDE cosmology constrained by Planck primaries. 
    Each step in the SA ladder is differentiated by color and by a vertical line. 
    The figure shows the same global optimizer run as in Fig. \ref{fig:global_min}, which also shows the temperature and jumping factors per SA step. 
    }
    \label{fig:lcdm_params_glob_min}
\end{figure*}

\begin{figure}
    \centering
    \includegraphics[width=0.45\textwidth]{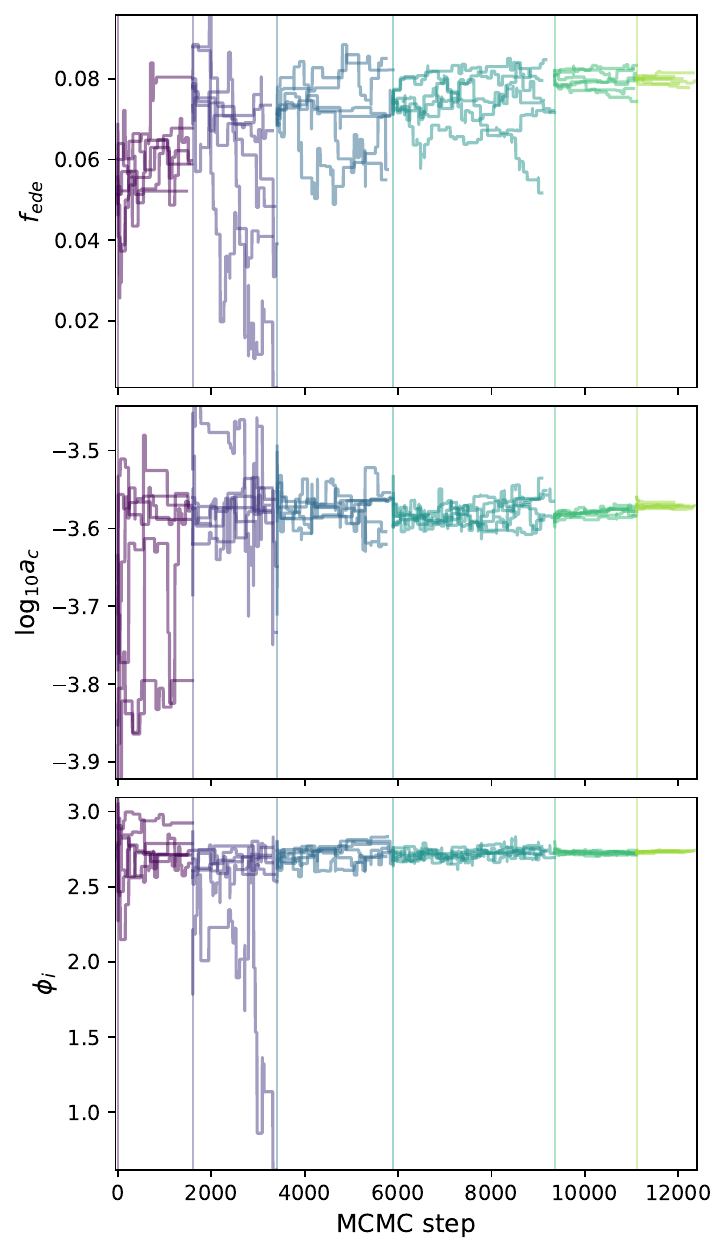}
    \caption{
    Convergence of EDE parameters over the same global optimizer run as in Figs. \ref{fig:global_min} and \ref{fig:lcdm_params_glob_min}. 
    Each step in the SA ladder is differentiated by color and by a vertical line. 
    The EDE parameters are the amount $\fede$ of EDE, the injection scale factor $\log_{10}a_c$ and the inital value $\phi_i$ of the scalar field. 
    }
    \label{fig:ede_param_glob_min}
\end{figure}

For global optimizers, we show how Procoli explores the parameter space, converging over SA steps towards the global minimum $\chi^2$. 
Specifically in Fig. \ref{fig:lcdm_params_glob_min}, we show the six \LCDM\ parameters when fit to Planck CMB primaries with an EDE cosmology. 
The EDE parameters for this optimization are shown in Fig. \ref{fig:ede_param_glob_min}. 
The corresponding progression in $\chi^2$ is shown in the main text in Fig. \ref{fig:global_min}. 

\begin{figure}
    \centering
    \includegraphics[width=0.458\textwidth]{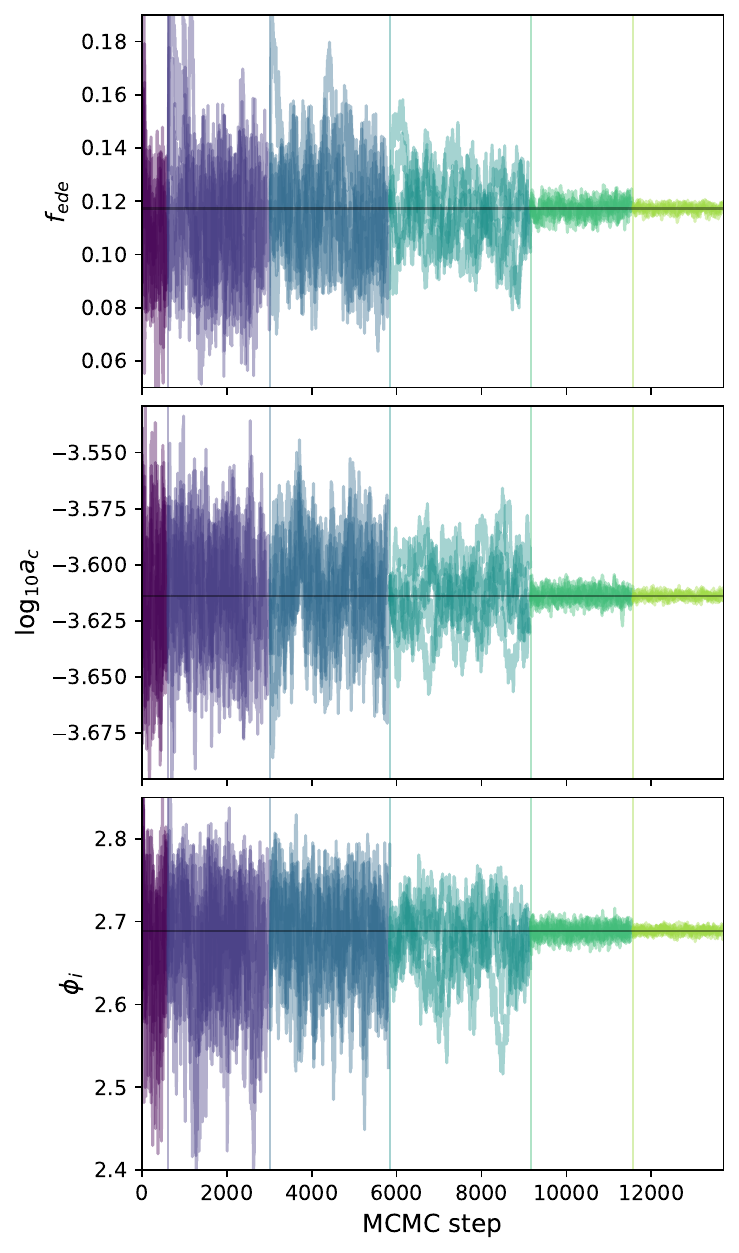}
    \caption{
    Convergence of EDE parameters to the fiducial cosmology in Table \ref{tab:ede_validation}, using Procoli default settings for the global optimization. 
    Each step in the SA ladder is differentiated by color and by a vertical line. 
    The horizontal black lines mark the positions of the input fiducial cosmology. 
    }
    \label{fig:ede_param_val}
\end{figure}

\begin{figure*}
    \centering
    \includegraphics[width=0.455\textwidth]{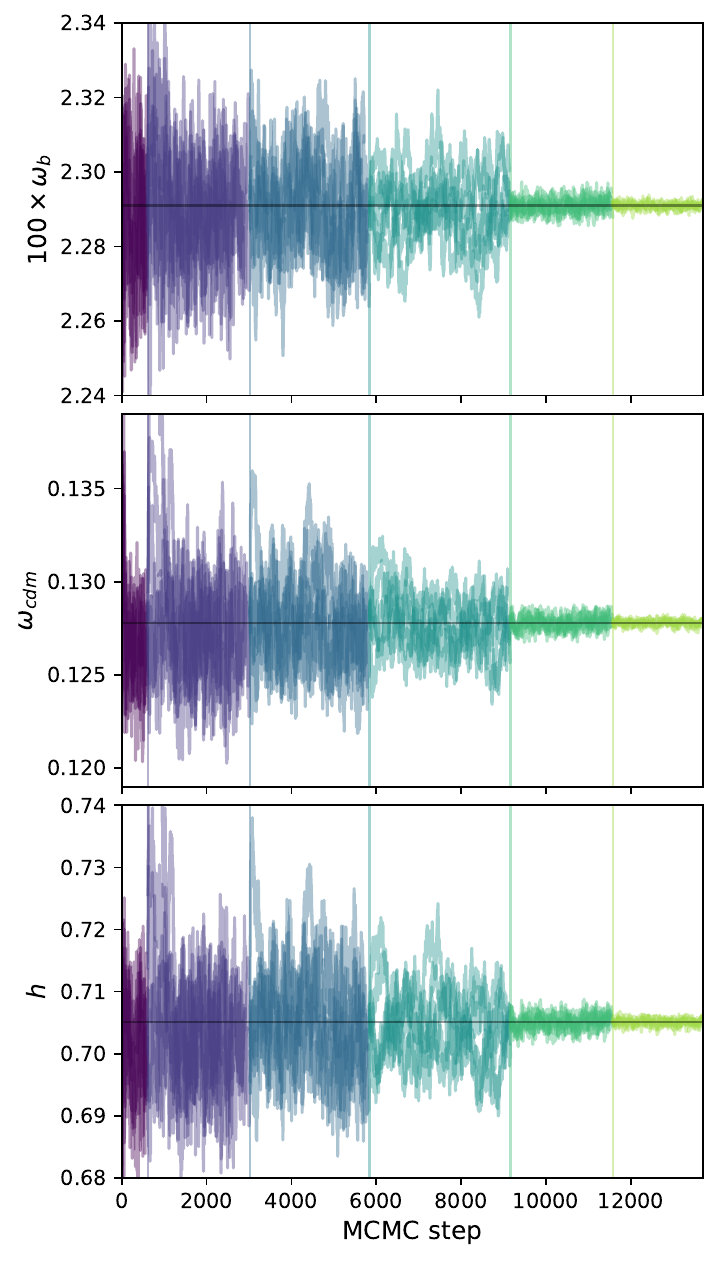}
    \includegraphics[width=0.45\textwidth]{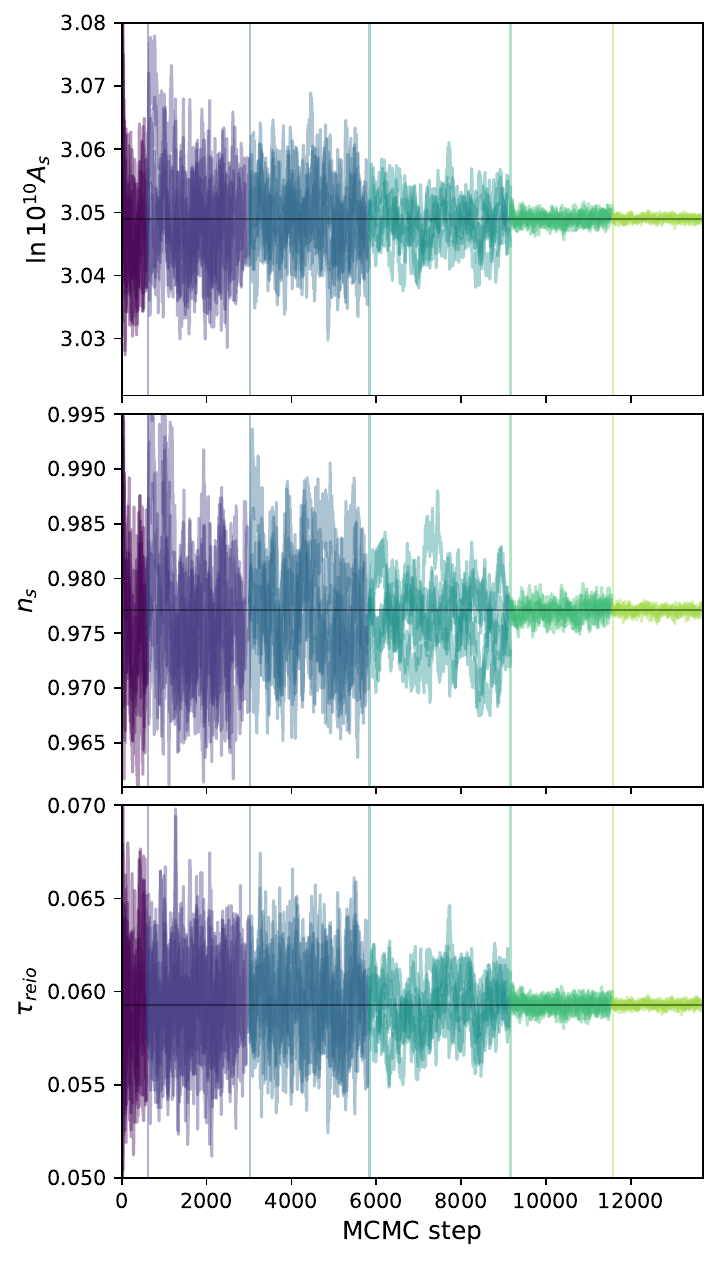}
    \caption{
    Convergence of the 6 \LCDM\ parameters to the fiducial cosmology in Table \ref{tab:ede_validation}, using Procoli default settings for the global optimization. 
    As before, each step in the SA ladder is differentiated by color and by a vertical line. 
    The horizontal black lines mark the positions of the input fiducial cosmology. 
    }
    \label{fig:lcdm_params_val}
\end{figure*}

Similarly, we also show the convergence of the optimizer to the fiducial cosmology for our validation tests in Figs. \ref{fig:ede_param_val} and \ref{fig:lcdm_params_val}. 
The true values of all nine cosmological parameters are correctly identified by Procoli. 
Compared to the curves in Figs. \ref{fig:lcdm_params_glob_min} and \ref{fig:ede_param_glob_min}, these curves are far more concentrated around the true input parameter values as the mock data generated was noiseless. 
The initial starting point is the scenario with $\chi^2 = 1162.78$ in Table \ref{tab:ede_validation}, which may be beyond the $y$-limits of the zoomed-in figures.

\appendix

\bibliography{paper}

\begin{thebibliography}{67}%
\makeatletter
\providecommand \@ifxundefined [1]{%
 \@ifx{#1\undefined}
}%
\providecommand \@ifnum [1]{%
 \ifnum #1\expandafter \@firstoftwo
 \else \expandafter \@secondoftwo
 \fi
}%
\providecommand \@ifx [1]{%
 \ifx #1\expandafter \@firstoftwo
 \else \expandafter \@secondoftwo
 \fi
}%
\providecommand \natexlab [1]{#1}%
\providecommand \enquote  [1]{``#1''}%
\providecommand \bibnamefont  [1]{#1}%
\providecommand \bibfnamefont [1]{#1}%
\providecommand \citenamefont [1]{#1}%
\providecommand \href@noop [0]{\@secondoftwo}%
\providecommand \href [0]{\begingroup \@sanitize@url \@href}%
\providecommand \@href[1]{\@@startlink{#1}\@@href}%
\providecommand \@@href[1]{\endgroup#1\@@endlink}%
\providecommand \@sanitize@url [0]{\catcode `\\12\catcode `\$12\catcode
  `\&12\catcode `\#12\catcode `\^12\catcode `\_12\catcode `\%12\relax}%
\providecommand \@@startlink[1]{}%
\providecommand \@@endlink[0]{}%
\providecommand \url  [0]{\begingroup\@sanitize@url \@url }%
\providecommand \@url [1]{\endgroup\@href {#1}{\urlprefix }}%
\providecommand \urlprefix  [0]{URL }%
\providecommand \Eprint [0]{\href }%
\providecommand \doibase [0]{http://dx.doi.org/}%
\providecommand \selectlanguage [0]{\@gobble}%
\providecommand \bibinfo  [0]{\@secondoftwo}%
\providecommand \bibfield  [0]{\@secondoftwo}%
\providecommand \translation [1]{[#1]}%
\providecommand \BibitemOpen [0]{}%
\providecommand \bibitemStop [0]{}%
\providecommand \bibitemNoStop [0]{.\EOS\space}%
\providecommand \EOS [0]{\spacefactor3000\relax}%
\providecommand \BibitemShut  [1]{\csname bibitem#1\endcsname}%
\let\auto@bib@innerbib\@empty
\bibitem [{\citenamefont {Lewis}\ and\ \citenamefont
  {Bridle}(2002)}]{Lewis:2002ah}%
  \BibitemOpen
  \bibfield  {author} {\bibinfo {author} {\bibfnamefont {Antony}\ \bibnamefont
  {Lewis}}\ and\ \bibinfo {author} {\bibfnamefont {Sarah}\ \bibnamefont
  {Bridle}},\ }\bibfield  {title} {\enquote {\bibinfo {title} {{Cosmological
  parameters from CMB and other data: A Monte Carlo approach}},}\ }\href
  {\doibase 10.1103/PhysRevD.66.103511} {\bibfield  {journal} {\bibinfo
  {journal} {Phys. Rev. D}\ }\textbf {\bibinfo {volume} {66}},\ \bibinfo
  {pages} {103511} (\bibinfo {year} {2002})},\ \Eprint
  {http://arxiv.org/abs/astro-ph/0205436} {arXiv:astro-ph/0205436} \BibitemShut
  {NoStop}%
\bibitem [{\citenamefont {Cousins}(1995)}]{Cousins:1994yw}%
  \BibitemOpen
  \bibfield  {author} {\bibinfo {author} {\bibfnamefont {Robert~D.}\
  \bibnamefont {Cousins}},\ }\bibfield  {title} {\enquote {\bibinfo {title}
  {{Why isn't every physicist a Bayesian?}}}\ }\href {\doibase 10.1119/1.17901}
  {\bibfield  {journal} {\bibinfo  {journal} {Am. J. Phys.}\ }\textbf {\bibinfo
  {volume} {63}},\ \bibinfo {pages} {398} (\bibinfo {year} {1995})}\BibitemShut
  {NoStop}%
\bibitem [{\citenamefont {Jeffreys}(1946)}]{Jeffreys:1946}%
  \BibitemOpen
  \bibfield  {author} {\bibinfo {author} {\bibfnamefont {Harold}\ \bibnamefont
  {Jeffreys}},\ }\bibfield  {title} {\enquote {\bibinfo {title} {{An invariant
  form for the prior probability in estimation problems}},}\ }\href {\doibase
  10.1098/rspa.1946.0056} {\bibfield  {journal} {\bibinfo  {journal} {Proc. R.
  Soc. A}\ }\textbf {\bibinfo {volume} {186}},\ \bibinfo {pages} {453–461}
  (\bibinfo {year} {1946})}\BibitemShut {NoStop}%
\bibitem [{\citenamefont {Feldman}\ and\ \citenamefont
  {Cousins}(1998)}]{Feldman:1997qc}%
  \BibitemOpen
  \bibfield  {author} {\bibinfo {author} {\bibfnamefont {Gary~J.}\ \bibnamefont
  {Feldman}}\ and\ \bibinfo {author} {\bibfnamefont {Robert~D.}\ \bibnamefont
  {Cousins}},\ }\bibfield  {title} {\enquote {\bibinfo {title} {{A Unified
  approach to the classical statistical analysis of small signals}},}\ }\href
  {\doibase 10.1103/PhysRevD.57.3873} {\bibfield  {journal} {\bibinfo
  {journal} {Phys. Rev. D}\ }\textbf {\bibinfo {volume} {57}},\ \bibinfo
  {pages} {3873--3889} (\bibinfo {year} {1998})},\ \Eprint
  {http://arxiv.org/abs/physics/9711021} {arXiv:physics/9711021} \BibitemShut
  {NoStop}%
\bibitem [{\citenamefont {Hill}\ \emph {et~al.}(2020)\citenamefont {Hill},
  \citenamefont {McDonough}, \citenamefont {Toomey},\ and\ \citenamefont
  {Alexander}}]{Hill:2020osr}%
  \BibitemOpen
  \bibfield  {author} {\bibinfo {author} {\bibfnamefont {J.~Colin}\
  \bibnamefont {Hill}}, \bibinfo {author} {\bibfnamefont {Evan}\ \bibnamefont
  {McDonough}}, \bibinfo {author} {\bibfnamefont {Michael~W.}\ \bibnamefont
  {Toomey}}, \ and\ \bibinfo {author} {\bibfnamefont {Stephon}\ \bibnamefont
  {Alexander}},\ }\bibfield  {title} {\enquote {\bibinfo {title} {{Early dark
  energy does not restore cosmological concordance}},}\ }\href {\doibase
  10.1103/PhysRevD.102.043507} {\bibfield  {journal} {\bibinfo  {journal}
  {Phys. Rev. D}\ }\textbf {\bibinfo {volume} {102}},\ \bibinfo {pages}
  {043507} (\bibinfo {year} {2020})},\ \Eprint
  {http://arxiv.org/abs/2003.07355} {arXiv:2003.07355 [astro-ph.CO]}
  \BibitemShut {NoStop}%
\bibitem [{\citenamefont {Ade}\ \emph {et~al.}(2014)\citenamefont {Ade} \emph
  {et~al.}}]{Planck:2013nga}%
  \BibitemOpen
  \bibfield  {author} {\bibinfo {author} {\bibfnamefont {P.~A.~R.}\
  \bibnamefont {Ade}} \emph {et~al.} (\bibinfo {collaboration} {Planck}),\
  }\bibfield  {title} {\enquote {\bibinfo {title} {{Planck intermediate
  results. XVI. Profile likelihoods for cosmological parameters}},}\ }\href
  {\doibase 10.1051/0004-6361/201323003} {\bibfield  {journal} {\bibinfo
  {journal} {Astron. Astrophys.}\ }\textbf {\bibinfo {volume} {566}},\ \bibinfo
  {pages} {A54} (\bibinfo {year} {2014})},\ \Eprint
  {http://arxiv.org/abs/1311.1657} {arXiv:1311.1657 [astro-ph.CO]} \BibitemShut
  {NoStop}%
\bibitem [{\citenamefont {Aghanim}\ \emph
  {et~al.}(2020{\natexlab{a}})\citenamefont {Aghanim} \emph
  {et~al.}}]{Planck:2018vyg}%
  \BibitemOpen
  \bibfield  {author} {\bibinfo {author} {\bibfnamefont {N.}~\bibnamefont
  {Aghanim}} \emph {et~al.} (\bibinfo {collaboration} {Planck}),\ }\bibfield
  {title} {\enquote {\bibinfo {title} {{Planck 2018 results. VI. Cosmological
  parameters}},}\ }\href {\doibase 10.1051/0004-6361/201833910} {\bibfield
  {journal} {\bibinfo  {journal} {Astron. Astrophys.}\ }\textbf {\bibinfo
  {volume} {641}},\ \bibinfo {pages} {A6} (\bibinfo {year}
  {2020}{\natexlab{a}})},\ \bibinfo {note} {[Erratum: Astron.Astrophys. 652, C4
  (2021)]},\ \Eprint {http://arxiv.org/abs/1807.06209} {arXiv:1807.06209
  [astro-ph.CO]} \BibitemShut {NoStop}%
\bibitem [{\citenamefont {Alam}\ \emph {et~al.}(2017)\citenamefont {Alam} \emph
  {et~al.}}]{BOSS:2016wmc}%
  \BibitemOpen
  \bibfield  {author} {\bibinfo {author} {\bibfnamefont {Shadab}\ \bibnamefont
  {Alam}} \emph {et~al.} (\bibinfo {collaboration} {BOSS}),\ }\bibfield
  {title} {\enquote {\bibinfo {title} {{The clustering of galaxies in the
  completed SDSS-III Baryon Oscillation Spectroscopic Survey: cosmological
  analysis of the DR12 galaxy sample}},}\ }\href {\doibase
  10.1093/mnras/stx721} {\bibfield  {journal} {\bibinfo  {journal} {Mon. Not.
  Roy. Astron. Soc.}\ }\textbf {\bibinfo {volume} {470}},\ \bibinfo {pages}
  {2617--2652} (\bibinfo {year} {2017})},\ \Eprint
  {http://arxiv.org/abs/1607.03155} {arXiv:1607.03155 [astro-ph.CO]}
  \BibitemShut {NoStop}%
\bibitem [{\citenamefont {Scolnic}\ \emph {et~al.}(2018)\citenamefont {Scolnic}
  \emph {et~al.}}]{Pan-STARRS1:2017jku}%
  \BibitemOpen
  \bibfield  {author} {\bibinfo {author} {\bibfnamefont {D.~M.}\ \bibnamefont
  {Scolnic}} \emph {et~al.} (\bibinfo {collaboration} {Pan-STARRS1}),\
  }\bibfield  {title} {\enquote {\bibinfo {title} {{The Complete Light-curve
  Sample of Spectroscopically Confirmed SNe Ia from Pan-STARRS1 and
  Cosmological Constraints from the Combined Pantheon Sample}},}\ }\href
  {\doibase 10.3847/1538-4357/aab9bb} {\bibfield  {journal} {\bibinfo
  {journal} {Astrophys. J.}\ }\textbf {\bibinfo {volume} {859}},\ \bibinfo
  {pages} {101} (\bibinfo {year} {2018})},\ \Eprint
  {http://arxiv.org/abs/1710.00845} {arXiv:1710.00845 [astro-ph.CO]}
  \BibitemShut {NoStop}%
\bibitem [{\citenamefont {Verde}\ \emph {et~al.}(2019)\citenamefont {Verde},
  \citenamefont {Treu},\ and\ \citenamefont {Riess}}]{Verde:2019ivm}%
  \BibitemOpen
  \bibfield  {author} {\bibinfo {author} {\bibfnamefont {L.}~\bibnamefont
  {Verde}}, \bibinfo {author} {\bibfnamefont {T.}~\bibnamefont {Treu}}, \ and\
  \bibinfo {author} {\bibfnamefont {A.~G.}\ \bibnamefont {Riess}},\ }\bibfield
  {title} {\enquote {\bibinfo {title} {{Tensions between the Early and the Late
  Universe}},}\ }\href {\doibase 10.1038/s41550-019-0902-0} {\bibfield
  {journal} {\bibinfo  {journal} {Nature Astron.}\ }\textbf {\bibinfo {volume}
  {3}},\ \bibinfo {pages} {891} (\bibinfo {year} {2019})},\ \Eprint
  {http://arxiv.org/abs/1907.10625} {arXiv:1907.10625 [astro-ph.CO]}
  \BibitemShut {NoStop}%
\bibitem [{\citenamefont {Abdalla}\ \emph {et~al.}(2022)\citenamefont {Abdalla}
  \emph {et~al.}}]{Abdalla:2022yfr}%
  \BibitemOpen
  \bibfield  {author} {\bibinfo {author} {\bibfnamefont {Elcio}\ \bibnamefont
  {Abdalla}} \emph {et~al.},\ }\bibfield  {title} {\enquote {\bibinfo {title}
  {{Cosmology intertwined: A review of the particle physics, astrophysics, and
  cosmology associated with the cosmological tensions and anomalies}},}\ }\href
  {\doibase 10.1016/j.jheap.2022.04.002} {\bibfield  {journal} {\bibinfo
  {journal} {JHEAp}\ }\textbf {\bibinfo {volume} {34}},\ \bibinfo {pages}
  {49--211} (\bibinfo {year} {2022})},\ \Eprint
  {http://arxiv.org/abs/2203.06142} {arXiv:2203.06142 [astro-ph.CO]}
  \BibitemShut {NoStop}%
\bibitem [{\citenamefont {Riess}\ and\ \citenamefont
  {Breuval}(2023)}]{Riess:2023egm}%
  \BibitemOpen
  \bibfield  {author} {\bibinfo {author} {\bibfnamefont {Adam~G.}\ \bibnamefont
  {Riess}}\ and\ \bibinfo {author} {\bibfnamefont {Louise}\ \bibnamefont
  {Breuval}},\ }\bibfield  {title} {\enquote {\bibinfo {title} {{The Local
  Value of H$_0$}},}\ \ }(\bibinfo {year} {2023})\ \Eprint
  {http://arxiv.org/abs/2308.10954} {arXiv:2308.10954 [astro-ph.CO]}
  \BibitemShut {NoStop}%
\bibitem [{\citenamefont {Secco}\ \emph {et~al.}(2023)\citenamefont {Secco},
  \citenamefont {Karwal}, \citenamefont {Hu},\ and\ \citenamefont
  {Krause}}]{Secco:2022kqg}%
  \BibitemOpen
  \bibfield  {author} {\bibinfo {author} {\bibfnamefont {Lucas~F.}\
  \bibnamefont {Secco}}, \bibinfo {author} {\bibfnamefont {Tanvi}\ \bibnamefont
  {Karwal}}, \bibinfo {author} {\bibfnamefont {Wayne}\ \bibnamefont {Hu}}, \
  and\ \bibinfo {author} {\bibfnamefont {Elisabeth}\ \bibnamefont {Krause}},\
  }\bibfield  {title} {\enquote {\bibinfo {title} {{Role of the Hubble scale in
  the weak lensing versus CMB tension}},}\ }\href {\doibase
  10.1103/PhysRevD.107.083532} {\bibfield  {journal} {\bibinfo  {journal}
  {Phys. Rev. D}\ }\textbf {\bibinfo {volume} {107}},\ \bibinfo {pages}
  {083532} (\bibinfo {year} {2023})},\ \Eprint
  {http://arxiv.org/abs/2209.12997} {arXiv:2209.12997 [astro-ph.CO]}
  \BibitemShut {NoStop}%
\bibitem [{\citenamefont {Amon}\ \emph {et~al.}(2022)\citenamefont {Amon} \emph
  {et~al.}}]{DES:2021bvc}%
  \BibitemOpen
  \bibfield  {author} {\bibinfo {author} {\bibfnamefont {A.}~\bibnamefont
  {Amon}} \emph {et~al.} (\bibinfo {collaboration} {DES}),\ }\bibfield  {title}
  {\enquote {\bibinfo {title} {{Dark Energy Survey Year 3 results: Cosmology
  from cosmic shear and robustness to data calibration}},}\ }\href {\doibase
  10.1103/PhysRevD.105.023514} {\bibfield  {journal} {\bibinfo  {journal}
  {Phys. Rev. D}\ }\textbf {\bibinfo {volume} {105}},\ \bibinfo {pages}
  {023514} (\bibinfo {year} {2022})},\ \Eprint
  {http://arxiv.org/abs/2105.13543} {arXiv:2105.13543 [astro-ph.CO]}
  \BibitemShut {NoStop}%
\bibitem [{\citenamefont {Secco}\ \emph {et~al.}(2022)\citenamefont {Secco}
  \emph {et~al.}}]{DES:2021vln}%
  \BibitemOpen
  \bibfield  {author} {\bibinfo {author} {\bibfnamefont {L.~F.}\ \bibnamefont
  {Secco}} \emph {et~al.} (\bibinfo {collaboration} {DES}),\ }\bibfield
  {title} {\enquote {\bibinfo {title} {{Dark Energy Survey Year 3 results:
  Cosmology from cosmic shear and robustness to modeling uncertainty}},}\
  }\href {\doibase 10.1103/PhysRevD.105.023515} {\bibfield  {journal} {\bibinfo
   {journal} {Phys. Rev. D}\ }\textbf {\bibinfo {volume} {105}},\ \bibinfo
  {pages} {023515} (\bibinfo {year} {2022})},\ \Eprint
  {http://arxiv.org/abs/2105.13544} {arXiv:2105.13544 [astro-ph.CO]}
  \BibitemShut {NoStop}%
\bibitem [{\citenamefont {Li}\ \emph {et~al.}(2023{\natexlab{a}})\citenamefont
  {Li} \emph {et~al.}}]{Li:2023azi}%
  \BibitemOpen
  \bibfield  {author} {\bibinfo {author} {\bibfnamefont {Shun-Sheng}\
  \bibnamefont {Li}} \emph {et~al.},\ }\bibfield  {title} {\enquote {\bibinfo
  {title} {{KiDS-1000: Cosmology with improved cosmic shear measurements}},}\
  }\href {\doibase 10.1051/0004-6361/202347236} {\bibfield  {journal} {\bibinfo
   {journal} {Astron. Astrophys.}\ }\textbf {\bibinfo {volume} {679}},\
  \bibinfo {pages} {A133} (\bibinfo {year} {2023}{\natexlab{a}})},\ \Eprint
  {http://arxiv.org/abs/2306.11124} {arXiv:2306.11124 [astro-ph.CO]}
  \BibitemShut {NoStop}%
\bibitem [{\citenamefont {Dalal}\ \emph {et~al.}(2023)\citenamefont {Dalal}
  \emph {et~al.}}]{Dalal:2023olq}%
  \BibitemOpen
  \bibfield  {author} {\bibinfo {author} {\bibfnamefont {Roohi}\ \bibnamefont
  {Dalal}} \emph {et~al.},\ }\bibfield  {title} {\enquote {\bibinfo {title}
  {{Hyper Suprime-Cam Year 3 results: Cosmology from cosmic shear power
  spectra}},}\ }\href {\doibase 10.1103/PhysRevD.108.123519} {\bibfield
  {journal} {\bibinfo  {journal} {Phys. Rev. D}\ }\textbf {\bibinfo {volume}
  {108}},\ \bibinfo {pages} {123519} (\bibinfo {year} {2023})},\ \Eprint
  {http://arxiv.org/abs/2304.00701} {arXiv:2304.00701 [astro-ph.CO]}
  \BibitemShut {NoStop}%
\bibitem [{\citenamefont {Smith}\ \emph {et~al.}(2021)\citenamefont {Smith},
  \citenamefont {Poulin}, \citenamefont {Bernal}, \citenamefont {Boddy},
  \citenamefont {Kamionkowski},\ and\ \citenamefont {Murgia}}]{Smith:2020rxx}%
  \BibitemOpen
  \bibfield  {author} {\bibinfo {author} {\bibfnamefont {Tristan~L.}\
  \bibnamefont {Smith}}, \bibinfo {author} {\bibfnamefont {Vivian}\
  \bibnamefont {Poulin}}, \bibinfo {author} {\bibfnamefont {Jos\'e~Luis}\
  \bibnamefont {Bernal}}, \bibinfo {author} {\bibfnamefont {Kimberly~K.}\
  \bibnamefont {Boddy}}, \bibinfo {author} {\bibfnamefont {Marc}\ \bibnamefont
  {Kamionkowski}}, \ and\ \bibinfo {author} {\bibfnamefont {Riccardo}\
  \bibnamefont {Murgia}},\ }\bibfield  {title} {\enquote {\bibinfo {title}
  {{Early dark energy is not excluded by current large-scale structure
  data}},}\ }\href {\doibase 10.1103/PhysRevD.103.123542} {\bibfield  {journal}
  {\bibinfo  {journal} {Phys. Rev. D}\ }\textbf {\bibinfo {volume} {103}},\
  \bibinfo {pages} {123542} (\bibinfo {year} {2021})},\ \Eprint
  {http://arxiv.org/abs/2009.10740} {arXiv:2009.10740 [astro-ph.CO]}
  \BibitemShut {NoStop}%
\bibitem [{\citenamefont {Herold}\ \emph {et~al.}(2022)\citenamefont {Herold},
  \citenamefont {Ferreira},\ and\ \citenamefont {Komatsu}}]{Herold:2021ksg}%
  \BibitemOpen
  \bibfield  {author} {\bibinfo {author} {\bibfnamefont {Laura}\ \bibnamefont
  {Herold}}, \bibinfo {author} {\bibfnamefont {Elisa G.~M.}\ \bibnamefont
  {Ferreira}}, \ and\ \bibinfo {author} {\bibfnamefont {Eiichiro}\ \bibnamefont
  {Komatsu}},\ }\bibfield  {title} {\enquote {\bibinfo {title} {{New Constraint
  on Early Dark Energy from Planck and BOSS Data Using the Profile
  Likelihood}},}\ }\href {\doibase 10.3847/2041-8213/ac63a3} {\bibfield
  {journal} {\bibinfo  {journal} {Astrophys. J. Lett.}\ }\textbf {\bibinfo
  {volume} {929}},\ \bibinfo {pages} {L16} (\bibinfo {year} {2022})},\ \Eprint
  {http://arxiv.org/abs/2112.12140} {arXiv:2112.12140 [astro-ph.CO]}
  \BibitemShut {NoStop}%
\bibitem [{\citenamefont {Hamann}(2012)}]{Hamann:2011hu}%
  \BibitemOpen
  \bibfield  {author} {\bibinfo {author} {\bibfnamefont {Jan}\ \bibnamefont
  {Hamann}},\ }\bibfield  {title} {\enquote {\bibinfo {title} {{Evidence for
  extra radiation? Profile likelihood versus Bayesian posterior}},}\ }\href
  {\doibase 10.1088/1475-7516/2012/03/021} {\bibfield  {journal} {\bibinfo
  {journal} {JCAP}\ }\textbf {\bibinfo {volume} {03}},\ \bibinfo {pages} {021}
  (\bibinfo {year} {2012})},\ \Eprint {http://arxiv.org/abs/1110.4271}
  {arXiv:1110.4271 [astro-ph.CO]} \BibitemShut {NoStop}%
\bibitem [{\citenamefont {Palanque-Delabrouille}\ \emph
  {et~al.}(2020)\citenamefont {Palanque-Delabrouille}, \citenamefont {Y\`eche},
  \citenamefont {Sch\"oneberg}, \citenamefont {Lesgourgues}, \citenamefont
  {Walther}, \citenamefont {Chabanier},\ and\ \citenamefont
  {Armengaud}}]{Palanque-Delabrouille:2019iyz}%
  \BibitemOpen
  \bibfield  {author} {\bibinfo {author} {\bibfnamefont {Nathalie}\
  \bibnamefont {Palanque-Delabrouille}}, \bibinfo {author} {\bibfnamefont
  {Christophe}\ \bibnamefont {Y\`eche}}, \bibinfo {author} {\bibfnamefont
  {Nils}\ \bibnamefont {Sch\"oneberg}}, \bibinfo {author} {\bibfnamefont
  {Julien}\ \bibnamefont {Lesgourgues}}, \bibinfo {author} {\bibfnamefont
  {Michael}\ \bibnamefont {Walther}}, \bibinfo {author} {\bibfnamefont
  {Sol\`ene}\ \bibnamefont {Chabanier}}, \ and\ \bibinfo {author}
  {\bibfnamefont {Eric}\ \bibnamefont {Armengaud}},\ }\bibfield  {title}
  {\enquote {\bibinfo {title} {{Hints, neutrino bounds and WDM constraints from
  SDSS DR14 Lyman-$\alpha$ and Planck full-survey data}},}\ }\href {\doibase
  10.1088/1475-7516/2020/04/038} {\bibfield  {journal} {\bibinfo  {journal}
  {JCAP}\ }\textbf {\bibinfo {volume} {04}},\ \bibinfo {pages} {038} (\bibinfo
  {year} {2020})},\ \Eprint {http://arxiv.org/abs/1911.09073} {arXiv:1911.09073
  [astro-ph.CO]} \BibitemShut {NoStop}%
\bibitem [{\citenamefont {Reeves}\ \emph {et~al.}(2023)\citenamefont {Reeves},
  \citenamefont {Herold}, \citenamefont {Vagnozzi}, \citenamefont {Sherwin},\
  and\ \citenamefont {Ferreira}}]{Reeves:2022aoi}%
  \BibitemOpen
  \bibfield  {author} {\bibinfo {author} {\bibfnamefont {Alexander}\
  \bibnamefont {Reeves}}, \bibinfo {author} {\bibfnamefont {Laura}\
  \bibnamefont {Herold}}, \bibinfo {author} {\bibfnamefont {Sunny}\
  \bibnamefont {Vagnozzi}}, \bibinfo {author} {\bibfnamefont {Blake~D.}\
  \bibnamefont {Sherwin}}, \ and\ \bibinfo {author} {\bibfnamefont {Elisa
  G.~M.}\ \bibnamefont {Ferreira}},\ }\bibfield  {title} {\enquote {\bibinfo
  {title} {{Restoring cosmological concordance with early dark energy and
  massive neutrinos?}}}\ }\href {\doibase 10.1093/mnras/stad317} {\bibfield
  {journal} {\bibinfo  {journal} {Mon. Not. Roy. Astron. Soc.}\ }\textbf
  {\bibinfo {volume} {520}},\ \bibinfo {pages} {3688--3695} (\bibinfo {year}
  {2023})},\ \Eprint {http://arxiv.org/abs/2207.01501} {arXiv:2207.01501
  [astro-ph.CO]} \BibitemShut {NoStop}%
\bibitem [{\citenamefont {Ranucci}(2012)}]{Ranucci:2012ed}%
  \BibitemOpen
  \bibfield  {author} {\bibinfo {author} {\bibfnamefont {Gioacchino}\
  \bibnamefont {Ranucci}},\ }\bibfield  {title} {\enquote {\bibinfo {title}
  {{The Profile likelihood ratio and the look elsewhere effect in high energy
  physics}},}\ }\href {\doibase 10.1016/j.nima.2011.09.047} {\bibfield
  {journal} {\bibinfo  {journal} {Nucl. Instrum. Meth. A}\ }\textbf {\bibinfo
  {volume} {661}},\ \bibinfo {pages} {77--85} (\bibinfo {year} {2012})},\
  \Eprint {http://arxiv.org/abs/1201.4604} {arXiv:1201.4604 [physics.data-an]}
  \BibitemShut {NoStop}%
\bibitem [{\citenamefont {Neyman}(1937)}]{Neyman:1937uhy}%
  \BibitemOpen
  \bibfield  {author} {\bibinfo {author} {\bibfnamefont {J.}~\bibnamefont
  {Neyman}},\ }\bibfield  {title} {\enquote {\bibinfo {title} {{Outline of a
  Theory of Statistical Estimation Based on the Classical Theory of
  Probability}},}\ }\href {\doibase 10.1098/rsta.1937.0005} {\bibfield
  {journal} {\bibinfo  {journal} {Phil. Trans. Roy. Soc. Lond. A}\ }\textbf
  {\bibinfo {volume} {236}},\ \bibinfo {pages} {333--380} (\bibinfo {year}
  {1937})}\BibitemShut {NoStop}%
\bibitem [{\citenamefont {Holm}\ \emph
  {et~al.}(2023{\natexlab{a}})\citenamefont {Holm}, \citenamefont {Nygaard},
  \citenamefont {Dakin}, \citenamefont {Hannestad},\ and\ \citenamefont
  {Tram}}]{Holm:2023uwa}%
  \BibitemOpen
  \bibfield  {author} {\bibinfo {author} {\bibfnamefont {Emil~Brinch}\
  \bibnamefont {Holm}}, \bibinfo {author} {\bibfnamefont {Andreas}\
  \bibnamefont {Nygaard}}, \bibinfo {author} {\bibfnamefont {Jeppe}\
  \bibnamefont {Dakin}}, \bibinfo {author} {\bibfnamefont {Steen}\ \bibnamefont
  {Hannestad}}, \ and\ \bibinfo {author} {\bibfnamefont {Thomas}\ \bibnamefont
  {Tram}},\ }\bibfield  {title} {\enquote {\bibinfo {title} {{PROSPECT: A
  profile likelihood code for frequentist cosmological parameter inference}},}\
  }\href@noop {} {\  (\bibinfo {year} {2023}{\natexlab{a}})},\ \Eprint
  {http://arxiv.org/abs/2312.02972} {arXiv:2312.02972 [astro-ph.CO]}
  \BibitemShut {NoStop}%
\bibitem [{\citenamefont {Nygaard}\ \emph {et~al.}(2023)\citenamefont
  {Nygaard}, \citenamefont {Holm}, \citenamefont {Hannestad},\ and\
  \citenamefont {Tram}}]{Nygaard:2022wri}%
  \BibitemOpen
  \bibfield  {author} {\bibinfo {author} {\bibfnamefont {Andreas}\ \bibnamefont
  {Nygaard}}, \bibinfo {author} {\bibfnamefont {Emil~Brinch}\ \bibnamefont
  {Holm}}, \bibinfo {author} {\bibfnamefont {Steen}\ \bibnamefont {Hannestad}},
  \ and\ \bibinfo {author} {\bibfnamefont {Thomas}\ \bibnamefont {Tram}},\
  }\bibfield  {title} {\enquote {\bibinfo {title} {{CONNECT: a neural network
  based framework for emulating cosmological observables and cosmological
  parameter inference}},}\ }\href {\doibase 10.1088/1475-7516/2023/05/025}
  {\bibfield  {journal} {\bibinfo  {journal} {JCAP}\ }\textbf {\bibinfo
  {volume} {05}},\ \bibinfo {pages} {025} (\bibinfo {year} {2023})},\ \Eprint
  {http://arxiv.org/abs/2205.15726} {arXiv:2205.15726 [astro-ph.IM]}
  \BibitemShut {NoStop}%
\bibitem [{\citenamefont {Li}\ \emph {et~al.}(2023{\natexlab{b}})\citenamefont
  {Li} \emph {et~al.}}]{Li:2022mdj}%
  \BibitemOpen
  \bibfield  {author} {\bibinfo {author} {\bibfnamefont {Zack}\ \bibnamefont
  {Li}} \emph {et~al.},\ }\bibfield  {title} {\enquote {\bibinfo {title} {{The
  Atacama Cosmology Telescope: limits on dark matter-baryon interactions from
  DR4 power spectra}},}\ }\href {\doibase 10.1088/1475-7516/2023/02/046}
  {\bibfield  {journal} {\bibinfo  {journal} {JCAP}\ }\textbf {\bibinfo
  {volume} {02}},\ \bibinfo {pages} {046} (\bibinfo {year}
  {2023}{\natexlab{b}})},\ \Eprint {http://arxiv.org/abs/2208.08985}
  {arXiv:2208.08985 [astro-ph.CO]} \BibitemShut {NoStop}%
\bibitem [{\citenamefont {Holm}\ \emph
  {et~al.}(2023{\natexlab{b}})\citenamefont {Holm}, \citenamefont {Herold},
  \citenamefont {Hannestad}, \citenamefont {Nygaard},\ and\ \citenamefont
  {Tram}}]{Holm:2022kkd}%
  \BibitemOpen
  \bibfield  {author} {\bibinfo {author} {\bibfnamefont {Emil~Brinch}\
  \bibnamefont {Holm}}, \bibinfo {author} {\bibfnamefont {Laura}\ \bibnamefont
  {Herold}}, \bibinfo {author} {\bibfnamefont {Steen}\ \bibnamefont
  {Hannestad}}, \bibinfo {author} {\bibfnamefont {Andreas}\ \bibnamefont
  {Nygaard}}, \ and\ \bibinfo {author} {\bibfnamefont {Thomas}\ \bibnamefont
  {Tram}},\ }\bibfield  {title} {\enquote {\bibinfo {title} {{Decaying dark
  matter with profile likelihoods}},}\ }\href {\doibase
  10.1103/PhysRevD.107.L021303} {\bibfield  {journal} {\bibinfo  {journal}
  {Phys. Rev. D}\ }\textbf {\bibinfo {volume} {107}},\ \bibinfo {pages}
  {L021303} (\bibinfo {year} {2023}{\natexlab{b}})},\ \Eprint
  {http://arxiv.org/abs/2211.01935} {arXiv:2211.01935 [astro-ph.CO]}
  \BibitemShut {NoStop}%
\bibitem [{\citenamefont {Audren}\ \emph {et~al.}(2013)\citenamefont {Audren},
  \citenamefont {Lesgourgues}, \citenamefont {Benabed},\ and\ \citenamefont
  {Prunet}}]{Audren:2012wb}%
  \BibitemOpen
  \bibfield  {author} {\bibinfo {author} {\bibfnamefont {Benjamin}\
  \bibnamefont {Audren}}, \bibinfo {author} {\bibfnamefont {Julien}\
  \bibnamefont {Lesgourgues}}, \bibinfo {author} {\bibfnamefont {Karim}\
  \bibnamefont {Benabed}}, \ and\ \bibinfo {author} {\bibfnamefont {Simon}\
  \bibnamefont {Prunet}},\ }\bibfield  {title} {\enquote {\bibinfo {title}
  {{Conservative Constraints on Early Cosmology: an illustration of the Monte
  Python cosmological parameter inference code}},}\ }\href {\doibase
  10.1088/1475-7516/2013/02/001} {\bibfield  {journal} {\bibinfo  {journal}
  {JCAP}\ }\textbf {\bibinfo {volume} {1302}},\ \bibinfo {pages} {001}
  (\bibinfo {year} {2013})},\ \Eprint {http://arxiv.org/abs/1210.7183}
  {arXiv:1210.7183 [astro-ph.CO]} \BibitemShut {NoStop}%
\bibitem [{\citenamefont {Brinckmann}\ and\ \citenamefont
  {Lesgourgues}(2019)}]{Brinckmann:2018cvx}%
  \BibitemOpen
  \bibfield  {author} {\bibinfo {author} {\bibfnamefont {Thejs}\ \bibnamefont
  {Brinckmann}}\ and\ \bibinfo {author} {\bibfnamefont {Julien}\ \bibnamefont
  {Lesgourgues}},\ }\bibfield  {title} {\enquote {\bibinfo {title}
  {{MontePython 3: boosted MCMC sampler and other features}},}\ }\href
  {\doibase 10.1016/j.dark.2018.100260} {\bibfield  {journal} {\bibinfo
  {journal} {Phys. Dark Univ.}\ }\textbf {\bibinfo {volume} {24}},\ \bibinfo
  {pages} {100260} (\bibinfo {year} {2019})},\ \Eprint
  {http://arxiv.org/abs/1804.07261} {arXiv:1804.07261 [astro-ph.CO]}
  \BibitemShut {NoStop}%
\bibitem [{\citenamefont {Blas}\ \emph {et~al.}(2011)\citenamefont {Blas},
  \citenamefont {Lesgourgues},\ and\ \citenamefont {Tram}}]{Blas:2011rf}%
  \BibitemOpen
  \bibfield  {author} {\bibinfo {author} {\bibfnamefont {Diego}\ \bibnamefont
  {Blas}}, \bibinfo {author} {\bibfnamefont {Julien}\ \bibnamefont
  {Lesgourgues}}, \ and\ \bibinfo {author} {\bibfnamefont {Thomas}\
  \bibnamefont {Tram}},\ }\bibfield  {title} {\enquote {\bibinfo {title} {{The
  Cosmic Linear Anisotropy Solving System (CLASS) II: Approximation
  schemes}},}\ }\href {\doibase 10.1088/1475-7516/2011/07/034} {\bibfield
  {journal} {\bibinfo  {journal} {JCAP}\ }\textbf {\bibinfo {volume} {07}},\
  \bibinfo {pages} {034} (\bibinfo {year} {2011})},\ \Eprint
  {http://arxiv.org/abs/1104.2933} {arXiv:1104.2933 [astro-ph.CO]} \BibitemShut
  {NoStop}%
\bibitem [{\citenamefont {Torrado}\ and\ \citenamefont
  {Lewis}(2021)}]{Torrado:2020dgo}%
  \BibitemOpen
  \bibfield  {author} {\bibinfo {author} {\bibfnamefont {Jesus}\ \bibnamefont
  {Torrado}}\ and\ \bibinfo {author} {\bibfnamefont {Antony}\ \bibnamefont
  {Lewis}},\ }\bibfield  {title} {\enquote {\bibinfo {title} {{Cobaya: Code for
  Bayesian Analysis of hierarchical physical models}},}\ }\href {\doibase
  10.1088/1475-7516/2021/05/057} {\bibfield  {journal} {\bibinfo  {journal}
  {JCAP}\ }\textbf {\bibinfo {volume} {05}},\ \bibinfo {pages} {057} (\bibinfo
  {year} {2021})},\ \Eprint {http://arxiv.org/abs/2005.05290} {arXiv:2005.05290
  [astro-ph.IM]} \BibitemShut {NoStop}%
\bibitem [{\citenamefont {{Torrado}}\ and\ \citenamefont
  {{Lewis}}(2019)}]{2019ascl.soft10019T}%
  \BibitemOpen
  \bibfield  {author} {\bibinfo {author} {\bibfnamefont {Jes{\'u}s}\
  \bibnamefont {{Torrado}}}\ and\ \bibinfo {author} {\bibfnamefont {Antony}\
  \bibnamefont {{Lewis}}},\ }\href@noop {} {\enquote {\bibinfo {title}
  {{Cobaya: Bayesian analysis in cosmology}},}\ }\bibinfo {howpublished}
  {Astrophysics Source Code Library, record ascl:1910.019} (\bibinfo {year}
  {2019}),\ \Eprint {http://arxiv.org/abs/1910.019} {ascl:1910.019}
  \BibitemShut {NoStop}%
\bibitem [{\citenamefont {Howlett}\ \emph {et~al.}(2012)\citenamefont
  {Howlett}, \citenamefont {Lewis}, \citenamefont {Hall},\ and\ \citenamefont
  {Challinor}}]{Howlett:2012mh}%
  \BibitemOpen
  \bibfield  {author} {\bibinfo {author} {\bibfnamefont {Cullan}\ \bibnamefont
  {Howlett}}, \bibinfo {author} {\bibfnamefont {Antony}\ \bibnamefont {Lewis}},
  \bibinfo {author} {\bibfnamefont {Alex}\ \bibnamefont {Hall}}, \ and\
  \bibinfo {author} {\bibfnamefont {Anthony}\ \bibnamefont {Challinor}},\
  }\bibfield  {title} {\enquote {\bibinfo {title} {{CMB power spectrum
  parameter degeneracies in the era of precision cosmology}},}\ }\href
  {\doibase 10.1088/1475-7516/2012/04/027} {\bibfield  {journal} {\bibinfo
  {journal} {JCAP}\ }\textbf {\bibinfo {volume} {04}},\ \bibinfo {pages} {027}
  (\bibinfo {year} {2012})},\ \Eprint {http://arxiv.org/abs/1201.3654}
  {arXiv:1201.3654 [astro-ph.CO]} \BibitemShut {NoStop}%
\bibitem [{\citenamefont {Lewis}\ \emph {et~al.}(2000)\citenamefont {Lewis},
  \citenamefont {Challinor},\ and\ \citenamefont {Lasenby}}]{Lewis:1999bs}%
  \BibitemOpen
  \bibfield  {author} {\bibinfo {author} {\bibfnamefont {Antony}\ \bibnamefont
  {Lewis}}, \bibinfo {author} {\bibfnamefont {Anthony}\ \bibnamefont
  {Challinor}}, \ and\ \bibinfo {author} {\bibfnamefont {Anthony}\ \bibnamefont
  {Lasenby}},\ }\bibfield  {title} {\enquote {\bibinfo {title} {{Efficient
  computation of CMB anisotropies in closed FRW models}},}\ }\href {\doibase
  10.1086/309179} {\bibfield  {journal} {\bibinfo  {journal} {Astrophys. J.}\
  }\textbf {\bibinfo {volume} {538}},\ \bibinfo {pages} {473--476} (\bibinfo
  {year} {2000})},\ \Eprint {http://arxiv.org/abs/astro-ph/9911177}
  {arXiv:astro-ph/9911177} \BibitemShut {NoStop}%
\bibitem [{\citenamefont {Hannestad}(2000)}]{Hannestad:2000wx}%
  \BibitemOpen
  \bibfield  {author} {\bibinfo {author} {\bibfnamefont {Steen}\ \bibnamefont
  {Hannestad}},\ }\bibfield  {title} {\enquote {\bibinfo {title} {{Stochastic
  optimization methods for extracting cosmological parameters from cosmic
  microwave background radiation power spectra}},}\ }\href {\doibase
  10.1103/PhysRevD.61.023002} {\bibfield  {journal} {\bibinfo  {journal} {Phys.
  Rev. D}\ }\textbf {\bibinfo {volume} {61}},\ \bibinfo {pages} {023002}
  (\bibinfo {year} {2000})},\ \Eprint {http://arxiv.org/abs/astro-ph/9911330}
  {arXiv:astro-ph/9911330} \BibitemShut {NoStop}%
\bibitem [{\citenamefont {Sch\"oneberg}\ \emph {et~al.}(2022)\citenamefont
  {Sch\"oneberg}, \citenamefont {Franco~Abell\'an}, \citenamefont
  {P\'erez~S\'anchez}, \citenamefont {Witte}, \citenamefont {Poulin},\ and\
  \citenamefont {Lesgourgues}}]{Schoneberg:2021qvd}%
  \BibitemOpen
  \bibfield  {author} {\bibinfo {author} {\bibfnamefont {Nils}\ \bibnamefont
  {Sch\"oneberg}}, \bibinfo {author} {\bibfnamefont {Guillermo}\ \bibnamefont
  {Franco~Abell\'an}}, \bibinfo {author} {\bibfnamefont {Andrea}\ \bibnamefont
  {P\'erez~S\'anchez}}, \bibinfo {author} {\bibfnamefont {Samuel~J.}\
  \bibnamefont {Witte}}, \bibinfo {author} {\bibfnamefont {Vivian}\
  \bibnamefont {Poulin}}, \ and\ \bibinfo {author} {\bibfnamefont {Julien}\
  \bibnamefont {Lesgourgues}},\ }\bibfield  {title} {\enquote {\bibinfo {title}
  {{The H0 Olympics: A fair ranking of proposed models}},}\ }\href {\doibase
  10.1016/j.physrep.2022.07.001} {\bibfield  {journal} {\bibinfo  {journal}
  {Phys. Rept.}\ }\textbf {\bibinfo {volume} {984}},\ \bibinfo {pages} {1--55}
  (\bibinfo {year} {2022})},\ \Eprint {http://arxiv.org/abs/2107.10291}
  {arXiv:2107.10291 [astro-ph.CO]} \BibitemShut {NoStop}%
\bibitem [{\citenamefont {Goldstein}\ \emph {et~al.}(2023)\citenamefont
  {Goldstein}, \citenamefont {Hill}, \citenamefont {Ir\v{s}i\v{c}},\ and\
  \citenamefont {Sherwin}}]{Goldstein:2023gnw}%
  \BibitemOpen
  \bibfield  {author} {\bibinfo {author} {\bibfnamefont {Samuel}\ \bibnamefont
  {Goldstein}}, \bibinfo {author} {\bibfnamefont {J.~Colin}\ \bibnamefont
  {Hill}}, \bibinfo {author} {\bibfnamefont {Vid}\ \bibnamefont
  {Ir\v{s}i\v{c}}}, \ and\ \bibinfo {author} {\bibfnamefont {Blake~D.}\
  \bibnamefont {Sherwin}},\ }\bibfield  {title} {\enquote {\bibinfo {title}
  {{Canonical Hubble-Tension-Resolving Early Dark Energy Cosmologies Are
  Inconsistent with the Lyman-\ensuremath{\alpha} Forest}},}\ }\href {\doibase
  10.1103/PhysRevLett.131.201001} {\bibfield  {journal} {\bibinfo  {journal}
  {Phys. Rev. Lett.}\ }\textbf {\bibinfo {volume} {131}},\ \bibinfo {pages}
  {201001} (\bibinfo {year} {2023})},\ \Eprint
  {http://arxiv.org/abs/2303.00746} {arXiv:2303.00746 [astro-ph.CO]}
  \BibitemShut {NoStop}%
\bibitem [{\citenamefont {Dunkley}\ \emph {et~al.}(2005)\citenamefont
  {Dunkley}, \citenamefont {Bucher}, \citenamefont {Ferreira}, \citenamefont
  {Moodley},\ and\ \citenamefont {Skordis}}]{Dunkley:2004sv}%
  \BibitemOpen
  \bibfield  {author} {\bibinfo {author} {\bibfnamefont {Joanna}\ \bibnamefont
  {Dunkley}}, \bibinfo {author} {\bibfnamefont {Martin}\ \bibnamefont
  {Bucher}}, \bibinfo {author} {\bibfnamefont {Pedro~G.}\ \bibnamefont
  {Ferreira}}, \bibinfo {author} {\bibfnamefont {Kavilan}\ \bibnamefont
  {Moodley}}, \ and\ \bibinfo {author} {\bibfnamefont {Constantinos}\
  \bibnamefont {Skordis}},\ }\bibfield  {title} {\enquote {\bibinfo {title}
  {{Fast and reliable mcmc for cosmological parameter estimation}},}\ }\href
  {\doibase 10.1111/j.1365-2966.2004.08464.x} {\bibfield  {journal} {\bibinfo
  {journal} {Mon. Not. Roy. Astron. Soc.}\ }\textbf {\bibinfo {volume} {356}},\
  \bibinfo {pages} {925--936} (\bibinfo {year} {2005})},\ \Eprint
  {http://arxiv.org/abs/astro-ph/0405462} {arXiv:astro-ph/0405462} \BibitemShut
  {NoStop}%
\bibitem [{\citenamefont {Dembinski}\ \emph {et~al.}(2023)\citenamefont
  {Dembinski}, \citenamefont {Ongmongkolkul}, \citenamefont {Deil},
  \citenamefont {Schreiner}, \citenamefont {Feickert}, \citenamefont {Burr},
  \citenamefont {Watson}, \citenamefont {Rost}, \citenamefont {Pearce},
  \citenamefont {Geiger}, \citenamefont {Abdelmotteleb}, \citenamefont {Desai},
  \citenamefont {Wiedemann}, \citenamefont {Gohlke}, \citenamefont {Sanders},
  \citenamefont {Drotleff}, \citenamefont {Eschle}, \citenamefont {Neste},
  \citenamefont {Gorelli}, \citenamefont {Baak}, \citenamefont {Eliachevitch},\
  and\ \citenamefont {Zapata}}]{dembinski_2023_8249703}%
  \BibitemOpen
  \bibfield  {author} {\bibinfo {author} {\bibfnamefont {Hans}\ \bibnamefont
  {Dembinski}}, \bibinfo {author} {\bibfnamefont {Piti}\ \bibnamefont
  {Ongmongkolkul}}, \bibinfo {author} {\bibfnamefont {Christoph}\ \bibnamefont
  {Deil}}, \bibinfo {author} {\bibfnamefont {Henry}\ \bibnamefont {Schreiner}},
  \bibinfo {author} {\bibfnamefont {Matthew}\ \bibnamefont {Feickert}},
  \bibinfo {author} {\bibfnamefont {Chris}\ \bibnamefont {Burr}}, \bibinfo
  {author} {\bibfnamefont {Jason}\ \bibnamefont {Watson}}, \bibinfo {author}
  {\bibfnamefont {Fabian}\ \bibnamefont {Rost}}, \bibinfo {author}
  {\bibfnamefont {Alex}\ \bibnamefont {Pearce}}, \bibinfo {author}
  {\bibfnamefont {Lukas}\ \bibnamefont {Geiger}}, \bibinfo {author}
  {\bibfnamefont {Ahmed}\ \bibnamefont {Abdelmotteleb}}, \bibinfo {author}
  {\bibfnamefont {Aman}\ \bibnamefont {Desai}}, \bibinfo {author}
  {\bibfnamefont {Bernhard~M.}\ \bibnamefont {Wiedemann}}, \bibinfo {author}
  {\bibfnamefont {Christoph}\ \bibnamefont {Gohlke}}, \bibinfo {author}
  {\bibfnamefont {Jeremy}\ \bibnamefont {Sanders}}, \bibinfo {author}
  {\bibfnamefont {Jonas}\ \bibnamefont {Drotleff}}, \bibinfo {author}
  {\bibfnamefont {Jonas}\ \bibnamefont {Eschle}}, \bibinfo {author}
  {\bibfnamefont {Ludwig}\ \bibnamefont {Neste}}, \bibinfo {author}
  {\bibfnamefont {Marco~Edward}\ \bibnamefont {Gorelli}}, \bibinfo {author}
  {\bibfnamefont {Max}\ \bibnamefont {Baak}}, \bibinfo {author} {\bibfnamefont
  {Michael}\ \bibnamefont {Eliachevitch}}, \ and\ \bibinfo {author}
  {\bibfnamefont {Omar}\ \bibnamefont {Zapata}},\ }\href {\doibase
  10.5281/zenodo.8249703} {\enquote {\bibinfo {title} {scikit-hep/iminuit},}\ }
  (\bibinfo {year} {2023})\BibitemShut {NoStop}%
\bibitem [{\citenamefont {James}\ and\ \citenamefont
  {Roos}(1975)}]{James:1975dr}%
  \BibitemOpen
  \bibfield  {author} {\bibinfo {author} {\bibfnamefont {F.}~\bibnamefont
  {James}}\ and\ \bibinfo {author} {\bibfnamefont {M.}~\bibnamefont {Roos}},\
  }\bibfield  {title} {\enquote {\bibinfo {title} {{Minuit: A System for
  Function Minimization and Analysis of the Parameter Errors and
  Correlations}},}\ }\href {\doibase 10.1016/0010-4655(75)90039-9} {\bibfield
  {journal} {\bibinfo  {journal} {Comput. Phys. Commun.}\ }\textbf {\bibinfo
  {volume} {10}},\ \bibinfo {pages} {343--367} (\bibinfo {year}
  {1975})}\BibitemShut {NoStop}%
\bibitem [{\citenamefont {Lewis}(2019)}]{Lewis:2019xzd}%
  \BibitemOpen
  \bibfield  {author} {\bibinfo {author} {\bibfnamefont {Antony}\ \bibnamefont
  {Lewis}},\ }\bibfield  {title} {\enquote {\bibinfo {title} {{GetDist: a
  Python package for analysing Monte Carlo samples}},}\ }\href@noop {} {\
  (\bibinfo {year} {2019})},\ \Eprint {http://arxiv.org/abs/1910.13970}
  {arXiv:1910.13970 [astro-ph.IM]} \BibitemShut {NoStop}%
\bibitem [{\citenamefont {Karwal}\ and\ \citenamefont
  {Kamionkowski}(2016)}]{Karwal:2016vyq}%
  \BibitemOpen
  \bibfield  {author} {\bibinfo {author} {\bibfnamefont {Tanvi}\ \bibnamefont
  {Karwal}}\ and\ \bibinfo {author} {\bibfnamefont {Marc}\ \bibnamefont
  {Kamionkowski}},\ }\bibfield  {title} {\enquote {\bibinfo {title} {{Dark
  energy at early times, the Hubble parameter, and the string axiverse}},}\
  }\href {\doibase 10.1103/PhysRevD.94.103523} {\bibfield  {journal} {\bibinfo
  {journal} {Phys. Rev. D}\ }\textbf {\bibinfo {volume} {94}},\ \bibinfo
  {pages} {103523} (\bibinfo {year} {2016})},\ \Eprint
  {http://arxiv.org/abs/1608.01309} {arXiv:1608.01309 [astro-ph.CO]}
  \BibitemShut {NoStop}%
\bibitem [{\citenamefont {Poulin}\ \emph {et~al.}(2019)\citenamefont {Poulin},
  \citenamefont {Smith}, \citenamefont {Karwal},\ and\ \citenamefont
  {Kamionkowski}}]{Poulin:2018cxd}%
  \BibitemOpen
  \bibfield  {author} {\bibinfo {author} {\bibfnamefont {Vivian}\ \bibnamefont
  {Poulin}}, \bibinfo {author} {\bibfnamefont {Tristan~L.}\ \bibnamefont
  {Smith}}, \bibinfo {author} {\bibfnamefont {Tanvi}\ \bibnamefont {Karwal}}, \
  and\ \bibinfo {author} {\bibfnamefont {Marc}\ \bibnamefont {Kamionkowski}},\
  }\bibfield  {title} {\enquote {\bibinfo {title} {{Early Dark Energy Can
  Resolve The Hubble Tension}},}\ }\href {\doibase
  10.1103/PhysRevLett.122.221301} {\bibfield  {journal} {\bibinfo  {journal}
  {Phys. Rev. Lett.}\ }\textbf {\bibinfo {volume} {122}},\ \bibinfo {pages}
  {221301} (\bibinfo {year} {2019})},\ \Eprint
  {http://arxiv.org/abs/1811.04083} {arXiv:1811.04083 [astro-ph.CO]}
  \BibitemShut {NoStop}%
\bibitem [{\citenamefont {Poulin}\ \emph {et~al.}(2018)\citenamefont {Poulin},
  \citenamefont {Smith}, \citenamefont {Grin}, \citenamefont {Karwal},\ and\
  \citenamefont {Kamionkowski}}]{Poulin:2018dzj}%
  \BibitemOpen
  \bibfield  {author} {\bibinfo {author} {\bibfnamefont {Vivian}\ \bibnamefont
  {Poulin}}, \bibinfo {author} {\bibfnamefont {Tristan~L.}\ \bibnamefont
  {Smith}}, \bibinfo {author} {\bibfnamefont {Daniel}\ \bibnamefont {Grin}},
  \bibinfo {author} {\bibfnamefont {Tanvi}\ \bibnamefont {Karwal}}, \ and\
  \bibinfo {author} {\bibfnamefont {Marc}\ \bibnamefont {Kamionkowski}},\
  }\bibfield  {title} {\enquote {\bibinfo {title} {{Cosmological implications
  of ultralight axionlike fields}},}\ }\href {\doibase
  10.1103/PhysRevD.98.083525} {\bibfield  {journal} {\bibinfo  {journal} {Phys.
  Rev. D}\ }\textbf {\bibinfo {volume} {98}},\ \bibinfo {pages} {083525}
  (\bibinfo {year} {2018})},\ \Eprint {http://arxiv.org/abs/1806.10608}
  {arXiv:1806.10608 [astro-ph.CO]} \BibitemShut {NoStop}%
\bibitem [{\citenamefont {Poulin}\ \emph {et~al.}(2023)\citenamefont {Poulin},
  \citenamefont {Smith},\ and\ \citenamefont {Karwal}}]{Poulin:2023lkg}%
  \BibitemOpen
  \bibfield  {author} {\bibinfo {author} {\bibfnamefont {Vivian}\ \bibnamefont
  {Poulin}}, \bibinfo {author} {\bibfnamefont {Tristan~L.}\ \bibnamefont
  {Smith}}, \ and\ \bibinfo {author} {\bibfnamefont {Tanvi}\ \bibnamefont
  {Karwal}},\ }\bibfield  {title} {\enquote {\bibinfo {title} {{The Ups and
  Downs of Early Dark Energy solutions to the Hubble tension: A review of
  models, hints and constraints circa 2023}},}\ }\href {\doibase
  10.1016/j.dark.2023.101348} {\bibfield  {journal} {\bibinfo  {journal} {Phys.
  Dark Univ.}\ }\textbf {\bibinfo {volume} {42}},\ \bibinfo {pages} {101348}
  (\bibinfo {year} {2023})},\ \Eprint {http://arxiv.org/abs/2302.09032}
  {arXiv:2302.09032 [astro-ph.CO]} \BibitemShut {NoStop}%
\bibitem [{\citenamefont {Kamionkowski}\ and\ \citenamefont
  {Riess}(2023)}]{Kamionkowski:2022pkx}%
  \BibitemOpen
  \bibfield  {author} {\bibinfo {author} {\bibfnamefont {Marc}\ \bibnamefont
  {Kamionkowski}}\ and\ \bibinfo {author} {\bibfnamefont {Adam~G.}\
  \bibnamefont {Riess}},\ }\bibfield  {title} {\enquote {\bibinfo {title} {{The
  Hubble Tension and Early Dark Energy}},}\ }\href@noop {} {\bibfield
  {journal} {\bibinfo  {journal} {Ann. Rev. Nucl. Part. Sci.}\ }\textbf
  {\bibinfo {volume} {73}},\ \bibinfo {pages} {153--180} (\bibinfo {year}
  {2023})},\ \Eprint {http://arxiv.org/abs/2211.04492} {arXiv:2211.04492
  [astro-ph.CO]} \BibitemShut {NoStop}%
\bibitem [{\citenamefont {Riess}\ \emph
  {et~al.}(2022{\natexlab{a}})\citenamefont {Riess}, \citenamefont {Breuval},
  \citenamefont {Yuan}, \citenamefont {Casertano}, \citenamefont {Macri},
  \citenamefont {Bowers}, \citenamefont {Scolnic}, \citenamefont
  {Cantat-Gaudin}, \citenamefont {Anderson},\ and\ \citenamefont
  {Reyes}}]{Riess:2022mme}%
  \BibitemOpen
  \bibfield  {author} {\bibinfo {author} {\bibfnamefont {Adam~G.}\ \bibnamefont
  {Riess}}, \bibinfo {author} {\bibfnamefont {Louise}\ \bibnamefont {Breuval}},
  \bibinfo {author} {\bibfnamefont {Wenlong}\ \bibnamefont {Yuan}}, \bibinfo
  {author} {\bibfnamefont {Stefano}\ \bibnamefont {Casertano}}, \bibinfo
  {author} {\bibfnamefont {Lucas~M.}\ \bibnamefont {Macri}}, \bibinfo {author}
  {\bibfnamefont {J.~Bradley}\ \bibnamefont {Bowers}}, \bibinfo {author}
  {\bibfnamefont {Dan}\ \bibnamefont {Scolnic}}, \bibinfo {author}
  {\bibfnamefont {Tristan}\ \bibnamefont {Cantat-Gaudin}}, \bibinfo {author}
  {\bibfnamefont {Richard~I.}\ \bibnamefont {Anderson}}, \ and\ \bibinfo
  {author} {\bibfnamefont {Mauricio~Cruz}\ \bibnamefont {Reyes}},\ }\bibfield
  {title} {\enquote {\bibinfo {title} {{Cluster Cepheids with High Precision
  Gaia Parallaxes, Low Zero-point Uncertainties, and Hubble Space Telescope
  Photometry}},}\ }\href {\doibase 10.3847/1538-4357/ac8f24} {\bibfield
  {journal} {\bibinfo  {journal} {Astrophys. J.}\ }\textbf {\bibinfo {volume}
  {938}},\ \bibinfo {pages} {36} (\bibinfo {year} {2022}{\natexlab{a}})},\
  \Eprint {http://arxiv.org/abs/2208.01045} {arXiv:2208.01045 [astro-ph.CO]}
  \BibitemShut {NoStop}%
\bibitem [{\citenamefont {Bernal}\ \emph {et~al.}(2016)\citenamefont {Bernal},
  \citenamefont {Verde},\ and\ \citenamefont {Riess}}]{Bernal:2016gxb}%
  \BibitemOpen
  \bibfield  {author} {\bibinfo {author} {\bibfnamefont {Jose~Luis}\
  \bibnamefont {Bernal}}, \bibinfo {author} {\bibfnamefont {Licia}\
  \bibnamefont {Verde}}, \ and\ \bibinfo {author} {\bibfnamefont {Adam~G.}\
  \bibnamefont {Riess}},\ }\bibfield  {title} {\enquote {\bibinfo {title} {{The
  trouble with $H_0$}},}\ }\href {\doibase 10.1088/1475-7516/2016/10/019}
  {\bibfield  {journal} {\bibinfo  {journal} {JCAP}\ }\textbf {\bibinfo
  {volume} {10}},\ \bibinfo {pages} {019} (\bibinfo {year} {2016})},\ \Eprint
  {http://arxiv.org/abs/1607.05617} {arXiv:1607.05617 [astro-ph.CO]}
  \BibitemShut {NoStop}%
\bibitem [{\citenamefont {Aylor}\ \emph {et~al.}(2019)\citenamefont {Aylor},
  \citenamefont {Joy}, \citenamefont {Knox}, \citenamefont {Millea},
  \citenamefont {Raghunathan},\ and\ \citenamefont {Wu}}]{Aylor:2018drw}%
  \BibitemOpen
  \bibfield  {author} {\bibinfo {author} {\bibfnamefont {Kevin}\ \bibnamefont
  {Aylor}}, \bibinfo {author} {\bibfnamefont {MacKenzie}\ \bibnamefont {Joy}},
  \bibinfo {author} {\bibfnamefont {Lloyd}\ \bibnamefont {Knox}}, \bibinfo
  {author} {\bibfnamefont {Marius}\ \bibnamefont {Millea}}, \bibinfo {author}
  {\bibfnamefont {Srinivasan}\ \bibnamefont {Raghunathan}}, \ and\ \bibinfo
  {author} {\bibfnamefont {W.~L.~Kimmy}\ \bibnamefont {Wu}},\ }\bibfield
  {title} {\enquote {\bibinfo {title} {{Sounds Discordant: Classical Distance
  Ladder \textbackslash{}\& $\Lambda$CDM -based Determinations of the
  Cosmological Sound Horizon}},}\ }\href {\doibase 10.3847/1538-4357/ab0898}
  {\bibfield  {journal} {\bibinfo  {journal} {Astrophys. J.}\ }\textbf
  {\bibinfo {volume} {874}},\ \bibinfo {pages} {4} (\bibinfo {year} {2019})},\
  \Eprint {http://arxiv.org/abs/1811.00537} {arXiv:1811.00537 [astro-ph.CO]}
  \BibitemShut {NoStop}%
\bibitem [{\citenamefont {Knox}\ and\ \citenamefont
  {Millea}(2020)}]{Knox:2019rjx}%
  \BibitemOpen
  \bibfield  {author} {\bibinfo {author} {\bibfnamefont {Lloyd}\ \bibnamefont
  {Knox}}\ and\ \bibinfo {author} {\bibfnamefont {Marius}\ \bibnamefont
  {Millea}},\ }\bibfield  {title} {\enquote {\bibinfo {title} {{Hubble constant
  hunter\textquoteright{}s guide}},}\ }\href {\doibase
  10.1103/PhysRevD.101.043533} {\bibfield  {journal} {\bibinfo  {journal}
  {Phys. Rev. D}\ }\textbf {\bibinfo {volume} {101}},\ \bibinfo {pages}
  {043533} (\bibinfo {year} {2020})},\ \Eprint
  {http://arxiv.org/abs/1908.03663} {arXiv:1908.03663 [astro-ph.CO]}
  \BibitemShut {NoStop}%
\bibitem [{\citenamefont {Lin}\ \emph {et~al.}(2019)\citenamefont {Lin},
  \citenamefont {Benevento}, \citenamefont {Hu},\ and\ \citenamefont
  {Raveri}}]{Lin:2019qug}%
  \BibitemOpen
  \bibfield  {author} {\bibinfo {author} {\bibfnamefont {Meng-Xiang}\
  \bibnamefont {Lin}}, \bibinfo {author} {\bibfnamefont {Giampaolo}\
  \bibnamefont {Benevento}}, \bibinfo {author} {\bibfnamefont {Wayne}\
  \bibnamefont {Hu}}, \ and\ \bibinfo {author} {\bibfnamefont {Marco}\
  \bibnamefont {Raveri}},\ }\bibfield  {title} {\enquote {\bibinfo {title}
  {{Acoustic Dark Energy: Potential Conversion of the Hubble Tension}},}\
  }\href {\doibase 10.1103/PhysRevD.100.063542} {\bibfield  {journal} {\bibinfo
   {journal} {Phys. Rev. D}\ }\textbf {\bibinfo {volume} {100}},\ \bibinfo
  {pages} {063542} (\bibinfo {year} {2019})},\ \Eprint
  {http://arxiv.org/abs/1905.12618} {arXiv:1905.12618 [astro-ph.CO]}
  \BibitemShut {NoStop}%
\bibitem [{\citenamefont {Agrawal}\ \emph {et~al.}(2023)\citenamefont
  {Agrawal}, \citenamefont {Cyr-Racine}, \citenamefont {Pinner},\ and\
  \citenamefont {Randall}}]{Agrawal:2019lmo}%
  \BibitemOpen
  \bibfield  {author} {\bibinfo {author} {\bibfnamefont {Prateek}\ \bibnamefont
  {Agrawal}}, \bibinfo {author} {\bibfnamefont {Francis-Yan}\ \bibnamefont
  {Cyr-Racine}}, \bibinfo {author} {\bibfnamefont {David}\ \bibnamefont
  {Pinner}}, \ and\ \bibinfo {author} {\bibfnamefont {Lisa}\ \bibnamefont
  {Randall}},\ }\bibfield  {title} {\enquote {\bibinfo {title} {{Rock
  \textquoteleft{}n\textquoteright{} roll solutions to the Hubble tension}},}\
  }\href {\doibase 10.1016/j.dark.2023.101347} {\bibfield  {journal} {\bibinfo
  {journal} {Phys. Dark Univ.}\ }\textbf {\bibinfo {volume} {42}},\ \bibinfo
  {pages} {101347} (\bibinfo {year} {2023})},\ \Eprint
  {http://arxiv.org/abs/1904.01016} {arXiv:1904.01016 [astro-ph.CO]}
  \BibitemShut {NoStop}%
\bibitem [{\citenamefont {Niedermann}\ and\ \citenamefont
  {Sloth}(2021)}]{Niedermann:2019olb}%
  \BibitemOpen
  \bibfield  {author} {\bibinfo {author} {\bibfnamefont {Florian}\ \bibnamefont
  {Niedermann}}\ and\ \bibinfo {author} {\bibfnamefont {Martin~S.}\
  \bibnamefont {Sloth}},\ }\bibfield  {title} {\enquote {\bibinfo {title} {{New
  early dark energy}},}\ }\href {\doibase 10.1103/PhysRevD.103.L041303}
  {\bibfield  {journal} {\bibinfo  {journal} {Phys. Rev. D}\ }\textbf {\bibinfo
  {volume} {103}},\ \bibinfo {pages} {L041303} (\bibinfo {year} {2021})},\
  \Eprint {http://arxiv.org/abs/1910.10739} {arXiv:1910.10739 [astro-ph.CO]}
  \BibitemShut {NoStop}%
\bibitem [{\citenamefont {Karwal}\ \emph {et~al.}(2022)\citenamefont {Karwal},
  \citenamefont {Raveri}, \citenamefont {Jain}, \citenamefont {Khoury},\ and\
  \citenamefont {Trodden}}]{Karwal:2021vpk}%
  \BibitemOpen
  \bibfield  {author} {\bibinfo {author} {\bibfnamefont {Tanvi}\ \bibnamefont
  {Karwal}}, \bibinfo {author} {\bibfnamefont {Marco}\ \bibnamefont {Raveri}},
  \bibinfo {author} {\bibfnamefont {Bhuvnesh}\ \bibnamefont {Jain}}, \bibinfo
  {author} {\bibfnamefont {Justin}\ \bibnamefont {Khoury}}, \ and\ \bibinfo
  {author} {\bibfnamefont {Mark}\ \bibnamefont {Trodden}},\ }\bibfield  {title}
  {\enquote {\bibinfo {title} {{Chameleon early dark energy and the Hubble
  tension}},}\ }\href {\doibase 10.1103/PhysRevD.105.063535} {\bibfield
  {journal} {\bibinfo  {journal} {Phys. Rev. D}\ }\textbf {\bibinfo {volume}
  {105}},\ \bibinfo {pages} {063535} (\bibinfo {year} {2022})},\ \Eprint
  {http://arxiv.org/abs/2106.13290} {arXiv:2106.13290 [astro-ph.CO]}
  \BibitemShut {NoStop}%
\bibitem [{\citenamefont {Berghaus}\ and\ \citenamefont
  {Karwal}(2020)}]{Berghaus:2019cls}%
  \BibitemOpen
  \bibfield  {author} {\bibinfo {author} {\bibfnamefont {Kim~V.}\ \bibnamefont
  {Berghaus}}\ and\ \bibinfo {author} {\bibfnamefont {Tanvi}\ \bibnamefont
  {Karwal}},\ }\bibfield  {title} {\enquote {\bibinfo {title} {{Thermal
  Friction as a Solution to the Hubble Tension}},}\ }\href {\doibase
  10.1103/PhysRevD.101.083537} {\bibfield  {journal} {\bibinfo  {journal}
  {Phys. Rev. D}\ }\textbf {\bibinfo {volume} {101}},\ \bibinfo {pages}
  {083537} (\bibinfo {year} {2020})},\ \Eprint
  {http://arxiv.org/abs/1911.06281} {arXiv:1911.06281 [astro-ph.CO]}
  \BibitemShut {NoStop}%
\bibitem [{\citenamefont {Berghaus}\ and\ \citenamefont
  {Karwal}(2023)}]{Berghaus:2022cwf}%
  \BibitemOpen
  \bibfield  {author} {\bibinfo {author} {\bibfnamefont {Kim~V.}\ \bibnamefont
  {Berghaus}}\ and\ \bibinfo {author} {\bibfnamefont {Tanvi}\ \bibnamefont
  {Karwal}},\ }\bibfield  {title} {\enquote {\bibinfo {title} {{Thermal
  friction as a solution to the Hubble and large-scale structure tensions}},}\
  }\href {\doibase 10.1103/PhysRevD.107.103515} {\bibfield  {journal} {\bibinfo
   {journal} {Phys. Rev. D}\ }\textbf {\bibinfo {volume} {107}},\ \bibinfo
  {pages} {103515} (\bibinfo {year} {2023})},\ \Eprint
  {http://arxiv.org/abs/2204.09133} {arXiv:2204.09133 [astro-ph.CO]}
  \BibitemShut {NoStop}%
\bibitem [{\citenamefont {McDonough}\ \emph {et~al.}(2022)\citenamefont
  {McDonough}, \citenamefont {Lin}, \citenamefont {Hill}, \citenamefont {Hu},\
  and\ \citenamefont {Zhou}}]{McDonough:2021pdg}%
  \BibitemOpen
  \bibfield  {author} {\bibinfo {author} {\bibfnamefont {Evan}\ \bibnamefont
  {McDonough}}, \bibinfo {author} {\bibfnamefont {Meng-Xiang}\ \bibnamefont
  {Lin}}, \bibinfo {author} {\bibfnamefont {J.~Colin}\ \bibnamefont {Hill}},
  \bibinfo {author} {\bibfnamefont {Wayne}\ \bibnamefont {Hu}}, \ and\ \bibinfo
  {author} {\bibfnamefont {Shengjia}\ \bibnamefont {Zhou}},\ }\bibfield
  {title} {\enquote {\bibinfo {title} {{Early dark sector, the Hubble tension,
  and the swampland}},}\ }\href {\doibase 10.1103/PhysRevD.106.043525}
  {\bibfield  {journal} {\bibinfo  {journal} {Phys. Rev. D}\ }\textbf {\bibinfo
  {volume} {106}},\ \bibinfo {pages} {043525} (\bibinfo {year} {2022})},\
  \Eprint {http://arxiv.org/abs/2112.09128} {arXiv:2112.09128 [astro-ph.CO]}
  \BibitemShut {NoStop}%
\bibitem [{\citenamefont {Brissenden}\ \emph {et~al.}(2024)\citenamefont
  {Brissenden}, \citenamefont {Dimopoulos},\ and\ \citenamefont
  {S\'anchez~L\'opez}}]{Brissenden:2023yko}%
  \BibitemOpen
  \bibfield  {author} {\bibinfo {author} {\bibfnamefont {Lucy}\ \bibnamefont
  {Brissenden}}, \bibinfo {author} {\bibfnamefont {Konstantinos}\ \bibnamefont
  {Dimopoulos}}, \ and\ \bibinfo {author} {\bibfnamefont {Samuel}\ \bibnamefont
  {S\'anchez~L\'opez}},\ }\bibfield  {title} {\enquote {\bibinfo {title}
  {{Non-oscillating early dark energy and quintessence from
  \ensuremath{\alpha}-attractors}},}\ }\href {\doibase
  10.1016/j.astropartphys.2024.102925} {\bibfield  {journal} {\bibinfo
  {journal} {Astropart. Phys.}\ }\textbf {\bibinfo {volume} {157}},\ \bibinfo
  {pages} {102925} (\bibinfo {year} {2024})},\ \Eprint
  {http://arxiv.org/abs/2301.03572} {arXiv:2301.03572 [astro-ph.CO]}
  \BibitemShut {NoStop}%
\bibitem [{\citenamefont {Braglia}\ \emph {et~al.}(2021)\citenamefont
  {Braglia}, \citenamefont {Ballardini}, \citenamefont {Finelli},\ and\
  \citenamefont {Koyama}}]{Braglia:2020auw}%
  \BibitemOpen
  \bibfield  {author} {\bibinfo {author} {\bibfnamefont {Matteo}\ \bibnamefont
  {Braglia}}, \bibinfo {author} {\bibfnamefont {Mario}\ \bibnamefont
  {Ballardini}}, \bibinfo {author} {\bibfnamefont {Fabio}\ \bibnamefont
  {Finelli}}, \ and\ \bibinfo {author} {\bibfnamefont {Kazuya}\ \bibnamefont
  {Koyama}},\ }\bibfield  {title} {\enquote {\bibinfo {title} {{Early modified
  gravity in light of the $H_0$ tension and LSS data}},}\ }\href {\doibase
  10.1103/PhysRevD.103.043528} {\bibfield  {journal} {\bibinfo  {journal}
  {Phys. Rev. D}\ }\textbf {\bibinfo {volume} {103}},\ \bibinfo {pages}
  {043528} (\bibinfo {year} {2021})},\ \Eprint
  {http://arxiv.org/abs/2011.12934} {arXiv:2011.12934 [astro-ph.CO]}
  \BibitemShut {NoStop}%
\bibitem [{\citenamefont {Gonzalez}\ \emph {et~al.}(2020)\citenamefont
  {Gonzalez}, \citenamefont {Hertzberg},\ and\ \citenamefont
  {Rompineve}}]{Gonzalez:2020fdy}%
  \BibitemOpen
  \bibfield  {author} {\bibinfo {author} {\bibfnamefont {Mark}\ \bibnamefont
  {Gonzalez}}, \bibinfo {author} {\bibfnamefont {Mark~P.}\ \bibnamefont
  {Hertzberg}}, \ and\ \bibinfo {author} {\bibfnamefont {Fabrizio}\
  \bibnamefont {Rompineve}},\ }\bibfield  {title} {\enquote {\bibinfo {title}
  {{Ultralight Scalar Decay and the Hubble Tension}},}\ }\href {\doibase
  10.1088/1475-7516/2020/10/028} {\bibfield  {journal} {\bibinfo  {journal}
  {JCAP}\ }\textbf {\bibinfo {volume} {10}},\ \bibinfo {pages} {028} (\bibinfo
  {year} {2020})},\ \Eprint {http://arxiv.org/abs/2006.13959} {arXiv:2006.13959
  [astro-ph.CO]} \BibitemShut {NoStop}%
\bibitem [{\citenamefont {Sakstein}\ and\ \citenamefont
  {Trodden}(2020)}]{Sakstein:2019fmf}%
  \BibitemOpen
  \bibfield  {author} {\bibinfo {author} {\bibfnamefont {Jeremy}\ \bibnamefont
  {Sakstein}}\ and\ \bibinfo {author} {\bibfnamefont {Mark}\ \bibnamefont
  {Trodden}},\ }\bibfield  {title} {\enquote {\bibinfo {title} {{Early Dark
  Energy from Massive Neutrinos as a Natural Resolution of the Hubble
  Tension}},}\ }\href {\doibase 10.1103/PhysRevLett.124.161301} {\bibfield
  {journal} {\bibinfo  {journal} {Phys. Rev. Lett.}\ }\textbf {\bibinfo
  {volume} {124}},\ \bibinfo {pages} {161301} (\bibinfo {year} {2020})},\
  \Eprint {http://arxiv.org/abs/1911.11760} {arXiv:1911.11760 [astro-ph.CO]}
  \BibitemShut {NoStop}%
\bibitem [{\citenamefont {Aghanim}\ \emph
  {et~al.}(2020{\natexlab{b}})\citenamefont {Aghanim} \emph
  {et~al.}}]{Planck:2019nip}%
  \BibitemOpen
  \bibfield  {author} {\bibinfo {author} {\bibfnamefont {N.}~\bibnamefont
  {Aghanim}} \emph {et~al.} (\bibinfo {collaboration} {Planck}),\ }\bibfield
  {title} {\enquote {\bibinfo {title} {{Planck 2018 results. V. CMB power
  spectra and likelihoods}},}\ }\href {\doibase 10.1051/0004-6361/201936386}
  {\bibfield  {journal} {\bibinfo  {journal} {Astron. Astrophys.}\ }\textbf
  {\bibinfo {volume} {641}},\ \bibinfo {pages} {A5} (\bibinfo {year}
  {2020}{\natexlab{b}})},\ \Eprint {http://arxiv.org/abs/1907.12875}
  {arXiv:1907.12875 [astro-ph.CO]} \BibitemShut {NoStop}%
\bibitem [{\citenamefont {Aghanim}\ \emph
  {et~al.}(2020{\natexlab{c}})\citenamefont {Aghanim} \emph
  {et~al.}}]{Planck:2018lbu}%
  \BibitemOpen
  \bibfield  {author} {\bibinfo {author} {\bibfnamefont {N.}~\bibnamefont
  {Aghanim}} \emph {et~al.} (\bibinfo {collaboration} {Planck}),\ }\bibfield
  {title} {\enquote {\bibinfo {title} {{Planck 2018 results. VIII.
  Gravitational lensing}},}\ }\href {\doibase 10.1051/0004-6361/201833886}
  {\bibfield  {journal} {\bibinfo  {journal} {Astron. Astrophys.}\ }\textbf
  {\bibinfo {volume} {641}},\ \bibinfo {pages} {A8} (\bibinfo {year}
  {2020}{\natexlab{c}})},\ \Eprint {http://arxiv.org/abs/1807.06210}
  {arXiv:1807.06210 [astro-ph.CO]} \BibitemShut {NoStop}%
\bibitem [{\citenamefont {Beutler}\ \emph {et~al.}(2011)\citenamefont
  {Beutler}, \citenamefont {Blake}, \citenamefont {Colless}, \citenamefont
  {Jones}, \citenamefont {Staveley-Smith}, \citenamefont {Campbell},
  \citenamefont {Parker}, \citenamefont {Saunders},\ and\ \citenamefont
  {Watson}}]{Beutler:2011hx}%
  \BibitemOpen
  \bibfield  {author} {\bibinfo {author} {\bibfnamefont {Florian}\ \bibnamefont
  {Beutler}}, \bibinfo {author} {\bibfnamefont {Chris}\ \bibnamefont {Blake}},
  \bibinfo {author} {\bibfnamefont {Matthew}\ \bibnamefont {Colless}}, \bibinfo
  {author} {\bibfnamefont {D.~Heath}\ \bibnamefont {Jones}}, \bibinfo {author}
  {\bibfnamefont {Lister}\ \bibnamefont {Staveley-Smith}}, \bibinfo {author}
  {\bibfnamefont {Lachlan}\ \bibnamefont {Campbell}}, \bibinfo {author}
  {\bibfnamefont {Quentin}\ \bibnamefont {Parker}}, \bibinfo {author}
  {\bibfnamefont {Will}\ \bibnamefont {Saunders}}, \ and\ \bibinfo {author}
  {\bibfnamefont {Fred}\ \bibnamefont {Watson}},\ }\bibfield  {title} {\enquote
  {\bibinfo {title} {{The 6dF Galaxy Survey: Baryon Acoustic Oscillations and
  the Local Hubble Constant}},}\ }\href {\doibase
  10.1111/j.1365-2966.2011.19250.x} {\bibfield  {journal} {\bibinfo  {journal}
  {Mon. Not. Roy. Astron. Soc.}\ }\textbf {\bibinfo {volume} {416}},\ \bibinfo
  {pages} {3017--3032} (\bibinfo {year} {2011})},\ \Eprint
  {http://arxiv.org/abs/1106.3366} {arXiv:1106.3366 [astro-ph.CO]} \BibitemShut
  {NoStop}%
\bibitem [{\citenamefont {Ross}\ \emph {et~al.}(2015)\citenamefont {Ross},
  \citenamefont {Samushia}, \citenamefont {Howlett}, \citenamefont {Percival},
  \citenamefont {Burden},\ and\ \citenamefont {Manera}}]{Ross:2014qpa}%
  \BibitemOpen
  \bibfield  {author} {\bibinfo {author} {\bibfnamefont {Ashley~J.}\
  \bibnamefont {Ross}}, \bibinfo {author} {\bibfnamefont {Lado}\ \bibnamefont
  {Samushia}}, \bibinfo {author} {\bibfnamefont {Cullan}\ \bibnamefont
  {Howlett}}, \bibinfo {author} {\bibfnamefont {Will~J.}\ \bibnamefont
  {Percival}}, \bibinfo {author} {\bibfnamefont {Angela}\ \bibnamefont
  {Burden}}, \ and\ \bibinfo {author} {\bibfnamefont {Marc}\ \bibnamefont
  {Manera}},\ }\bibfield  {title} {\enquote {\bibinfo {title} {{The clustering
  of the SDSS DR7 main Galaxy sample \textendash{} I. A 4 per cent distance
  measure at $z = 0.15$}},}\ }\href {\doibase 10.1093/mnras/stv154} {\bibfield
  {journal} {\bibinfo  {journal} {Mon. Not. Roy. Astron. Soc.}\ }\textbf
  {\bibinfo {volume} {449}},\ \bibinfo {pages} {835--847} (\bibinfo {year}
  {2015})},\ \Eprint {http://arxiv.org/abs/1409.3242} {arXiv:1409.3242
  [astro-ph.CO]} \BibitemShut {NoStop}%
\bibitem [{\citenamefont {Riess}\ \emph
  {et~al.}(2022{\natexlab{b}})\citenamefont {Riess} \emph
  {et~al.}}]{Riess:2021jrx}%
  \BibitemOpen
  \bibfield  {author} {\bibinfo {author} {\bibfnamefont {Adam~G.}\ \bibnamefont
  {Riess}} \emph {et~al.},\ }\bibfield  {title} {\enquote {\bibinfo {title} {{A
  Comprehensive Measurement of the Local Value of the Hubble Constant with 1 km
  s$^{−1}$ Mpc$^{−1}$ Uncertainty from the Hubble Space Telescope and the
  SH0ES Team}},}\ }\href {\doibase 10.3847/2041-8213/ac5c5b} {\bibfield
  {journal} {\bibinfo  {journal} {Astrophys. J. Lett.}\ }\textbf {\bibinfo
  {volume} {934}},\ \bibinfo {pages} {L7} (\bibinfo {year}
  {2022}{\natexlab{b}})},\ \Eprint {http://arxiv.org/abs/2112.04510}
  {arXiv:2112.04510 [astro-ph.CO]} \BibitemShut {NoStop}%
\end{thebibliography}%
\end{document}